\documentclass[letterpaper,12pt,leqno]{article}
\usepackage{amsmath}
\usepackage{paper}
\usepackage{graphicx}
\usepackage{subcaption}
\usepackage{algorithm2e} 
\usepackage{ulem} 
\usepackage{threeparttable} 

\usepackage{float}
\hypersetup{pdftitle={Paper Example}}

\newcommand{\unit}{\mathrm{UP}}
\newcommand{\ttime}{\mathrm{TP}}
\newcommand{\cond}{\mathrm{C}}
\newcommand{\marg}{\mathrm{M}}
\newcommand{\ystar}{{Y}_{\istar\tstar}}
\newcommand{\hystar}{\hat{Y}_{\istar\tstar}}

\newcommand{\hsu}{\hat{\sigma}^{2,\mathrm{\unit}}_{NT}}
\newcommand{\hst}{\hat{\sigma}^{2,\mathrm{\ttime}}_{NT}}
\newcommand{\hsm}{\hat{\sigma}^{2,\mathrm{\marg}}_{NT}}
\newcommand{\hsc}{\hat{\sigma}^{2,\mathrm{\cond}}_{NT}}

\newcommand{\tsm}{\tilde{\sigma}^{2,M}_{\istar\tstar}}
\newcommand{\tsc}{\tilde{\sigma}^{2,C}_{\istar\tstar}}

\newcommand{\tsuit}{\tilde{\sigma}^{2,UP}_{it}}
\newcommand{\tstit}{\tilde{\sigma}^{2,TP}_{it}}
\newcommand{\tsmit}{\tilde{\sigma}^{2,M}_{it}}
\newcommand{\tscit}{\tilde{\sigma}^{2,C}_{it}}

\renewcommand{\mathbf}{\boldsymbol}
\renewcommand{\appendix}{\footnotesize\parindent 0cm\setcounter{equation}{0}
	\renewcommand{\theequation}{A.\arabic{equation}}
	\setcounter{lemma}{0}\renewcommand{\thelemma}{A.\arabic{lemma}}}

\newcommand{\by}{{\mathbf{Y}}}

\newcommand{\bw}{{\mathbf{W}}}

\newcommand{\istar}{{i^*}}

\newcommand{\indep}{\perp\!\!\!\perp}

\newcommand{\mme}{{\mathbb{E}}}
\newcommand{\mmv}{{\mathbb{V}}}

\newcommand{\tstar}{{t^*}}

 \usepackage[colorinlistoftodos]{todonotes}

\newcommand{\bib}{refs.bib}

\begin{document}
\RestyleAlgo{ruled}

\title{Estimating Variances for Causal Panel Data Estimators}
\author{Alexander Almeida, Susan Athey, Guido Imbens, Eva Lestant \& Alexia Olaizola
\thanks{
Alexander Almeida, almeidaa@stanford.edu, Graduate School of Business, Stanford University;
Susan Athey, Graduate School of Business, Stanford University
Guido Imbens, Stanford University;
Eva Lestant, elestant@stanford.edu, Department of Economics, Stanford University;
Alexia Olaizola, aolaizola@stanford.edu, Department of Economics, Stanford University.
We are grateful for comments at a presentation at the ASSA meetings in January 2025.
This work was partially supported by the Office of Naval Research under grant
N00014-17-1-2131 and 
 N00014-19-1-2468 and a gift from Amazon. 
 JEL No. C01, C12, C23, C31, 
}}
\date{November 2025}              
\begin{titlepage}\maketitle 
There has been a recent surge in research on causal panel data models, leading to many new estimators for average causal effects.  However, researchers have paid less attention to quantifying the precision of these estimators. This paper addresses that gap by studying the problem of variance estimation in causal panel settings. We develop a unified framework for comparing the three main variance estimators used in these settings: regression-based, Unit-Placebo, and Time-Placebo estimators. We show that each relies on a distinct exchangeability assumption and, correspondingly, each targets a different conditional variance. We find that, under some assumptions, all three estimators are all valid, but that their statistical power differs substantially depending on the heteroskedasticity present in the data. Building on these insights, we propose a new variance estimator that flexibly accounts for heteroskedasticity across the unit and time dimensions, and delivers superior statistical power in realistic panel data settings.

\end{titlepage}

\newpage

\section{Introduction}\label{s:introduction}

We study variance estimation for a  class of causal estimators in panel data settings. We focus on a balanced panel setting where we observe outcomes $Y_{it}$ for $N$ units, $i=1,\ldots,N$, and for $T$ periods, $t=1,\ldots,T$. Some of the units are exposed in some periods  to a binary treatment, with the treatment denoted by  $W_{it}\in\{0,1\}$. Taking a potential outcome perspective 
\citep{rubin1974estimating, imbens2015causal}, and assuming there is no interference/spillovers ({\it i.e.,} assuming SUTVA holds,  \citealt{rubin1978bayesian}), we denote the pair of potential outcomes for unit $i$ and time $t$ by $(Y_{it}(0),Y_{it}(1))$, which are related to the realized/observed outcome through the equality
\begin{equation}Y_{it}\equiv Y_{it}(W_{it})=
\left\{\begin{array}{ll}
Y_{it}(1)\qquad & \mathrm{if}\  W_{it}=1,\\
Y_{it}(0) & \mathrm{otherwise}.\end{array}\right.\end{equation}
$\bw$ and $\by$ denote the $N\times T$ matrices with typical element $W_{it}$ and $Y_{it}$ respectively.

We focus primarily  on the case with just a single unit/period pair exposed to the treatment, although the results are relevant for the general case with more complex assignment patterns as well. Let $\istar$ and $\tstar$ denote the unit and period that is treated, so $W_{\istar\tstar}=1$, and $W_{it}=0$ for all $(i,t)\neq(\istar,\tstar)$.\footnote{In our examples, it will often be the case that $\istar = N$ and $\tstar=T$. However, to avoid confusion in situations where this is not the case, we use $(\istar, \tstar)$ throughout.} The object of interest is the causal effect $\tau\equiv \ystar(1)-\ystar(0)$. For the treated unit and period pair $(\istar,\tstar)$, we observe the potential outcome given the treatment, $\ystar(1)$, but do not observe the potential control outcome $\ystar(0)$. In order to estimate $\tau$, we need to impute the missing potential outcome $\ystar(0)$. Denote the imputed value by $\hystar(0)$ and the estimated treatment effect by $\hat\tau=\ystar(1)-\hat{Y}_{\istar\tstar}(0)=\ystar-\hat{Y}_{\istar\tstar}(0)$. The three questions we address in this paper are $(i)$ how to \textit{conceptualize} the uncertainty for generic estimators in this setting, $(ii)$ how to \textit{estimate} this uncertainty, and $(iii)$ how to \textit{choose} between different variance estimators in practice. 

In this specific setting, multiple estimators for the causal effect $\tau$ have been proposed. Among the estimators widely used in modern practice  are Difference-In-Differences  (DID) or Two-Way-Fixed-Effect (TWFE) estimators, ({\it e.g.,} \citealt{Bertrand2004did, angrist2008mostly}), factor model estimators \citep{xu2017generalized, athey2017matrix}, as well as various
synthetic control estimators \citep{abadie2003, abadie2010synthetic, abadie2015comparative, arkhangelsky2021synthetic, ben2021augmented, abadie2021using, ben2022synthetic}. Given an estimator, whether it is the original DID estimator or one of the newer factor model or  SC-type estimators, the question we focus on in this paper is how to quantify its uncertainty. The construction of confidence or prediction intervals is particularly challenging in the setting with only a single unit/period pair exposed to the treatment, because large sample approximations cannot be used to motivate normality for the distribution of the estimator for the causal effect. As a result, confidence intervals often rely on distributional assumptions. Nevertheless, in some cases, it is possible to estimate the variance of the estimator accurately under substantially weaker assumptions. There have been a number of distinct estimators proposed for this variance in the literature, including regression-based variance estimators and placebo-based methods, with their specific properties established under different sets of assumptions. 
Typically, these variance estimators have been proposed in conjunction with specific point estimators, {\it e.g.} regression-based estimators for difference-in-differences estimators, and unit and time-placebo variance estimators for synthetic control estimators. Here, like
\citet{chernozhukov2021exact}, we look at variance estimation for generic point estimators.

In our first contribution, we put the previously proposed variance estimators in a common framework. We use that framework to show how the variance estimators proposed in the literature are conceptually different, and can lead to substantially different values and properties in practice. Most of the previous literature has primarily focused on the validity of the various variance estimators. The literature offers limited guidance on the comparative advantages of different variance estimators regarding precision and power. In this paper, we clarify these issues and provide recommendations for selecting an appropriate variance estimator.  In our second contribution, we propose a new class of variance estimators with superior power properties. 

There are currently three general approaches in the literature to variance estimation for panel data with a single treated unit/period pair.\footnote{
We note that there also methods that directly focus on the construction of confidence or prediction intervals, {\it e.g.,}
\cite{lei2021conformal, cattaneo2021prediction, cattaneo2025uncertainty}.}
We slightly modify all three of the proposed variance estimators to characterize them as  averages of squared residuals, that is, as placebo variance estimators. To calculate the residuals that all three of these variance estimators are implicitly based on, the true treated unit/period pair and a control unit/period pair are set aside. The analyst's preferred estimator 
is used to estimate the control outcome $Y_{it}(0)$ for the control unit/time period pair. The difference 
$Y_{it}(0)-\hat{Y}_{it}(0)$ between the actual observed control outcome and the estimated control outcome for that unit/time period pair is used as an estimate of the residual which in turn form the basis for estimating the variance.  

The first of the three variance estimators that we consider  is common in the TWFE/DID literature. There, it can be interpreted as the conventional homoskedastic least squares variance estimator. We can rewrite this variance estimator as approximately equal to the average of all squared residuals. 
We refer to this as the {\it Marginal} (M) variance estimator.
In the SC literature, two alternatives to this least squares variance estimator have been proposed. Both of these alternatives, unlike the marginal estimator, are typically motivated by placebo approaches. 
One of the two alternatives only uses residuals from unit/time-period pairs $(i,\tstar)$, that is, from all control units in the same period $\tstar$ as the treated unit/period pair \citep{abadie2010synthetic, doudchenko2016balancing}. We refer to this variance estimator as the {\it Unit-Placebo} (UP) variance estimator since it {averages over control units} in the treated period. This unit-placebo variance estimator is approximately equal to the average squared residual, averaged now only over unit/period pairs in the treated period.
The other proposal only uses residuals from unit/time-period pairs $(\istar,t)$, that is, from the treated unit $\istar$ in all control time periods (see, e.g., \citealt{doudchenko2016balancing, chernozhukov2021exact}). 
We refer to this as the {\it Time-Placebo} (TP) variance estimator since it averages over control time periods for the treated unit.
This time-placebo variance estimator is approximately equal to the average squared residual, averaged now only over control periods for the treated unit.
In sum, we have three possible estimators that all use average squared residuals but over different sets of units and time periods: M, the entire matrix of squared residuals; UP, the squared residuals for the other control units in the treated period $\tstar$; TP, the squared residuals for the same unit $\istar$ at different time periods.

If there is no heteroskedasticity of any type, the marginal (M) variance estimator is the most attractive.
We show that the unit-placebo (UP) variance estimator is relatively attractive in settings with heteroskedasticity in the error terms over time.
In contrast, the time-placebo (TP) variance estimator is attractive in settings with heteroskedasticity across units.
We then introduce a novel variance estimator that accounts for heteroskedasticity across both units and time periods. We refer to this new variance estimator as the {\it Conditional} (C) variance estimator. We provide details on this variance estimator in Section \ref{section:new estimator}.

As a motivating example for the questions we discuss in this paper, we estimate treatment effects and variances for three panel datasets widely used in the recent panel data literature: the California smoking data \citep{abadie2010synthetic}, the German unification data \citep{abadie2015comparative}, and the Mariel Boatlift data \citep{cardmariel, peri2019labor}.
In Table \ref{tab:motivating_differences}, we consider a single treated unit/time period cell, which is the actual treated unit in the first period it was treated ({\it i.e.,} California in 1989 for the smoking data, Germany in 1990 in the German Unification data, and Miami in 1980 in the Mariel Boatlift data). For this unit/period, we estimate the treatment effect using the SC estimator (similar results for the TWFE and SDID estimators are reported in the Appendix  \ref{tab_app:table1_twfe} for TWFE and \ref{tab_app:table1_sdid} for SDID).
We then compute the standard error using the square root of four variance estimators (the three, M, UP, and TP introduced so far, and the new one, C, for Conditional, to be formally defined and discussed in Section \ref{section:new estimator}).

The key finding in this table is that the resulting standard errors differ substantially across variance estimators. For example, in the California Smoking dataset, the M standard error is almost nine times as large as the C  standard error. In the German unification data, the UP  standard error  is more than ten times as large as the TP standard error. The important point is not which estimator is larger in any given case--this can vary depending on the application--but rather that the differences across estimators are sometimes an order of magnitude or more.  
These findings raise the main questions addressed in this paper: namely, what causes these differences? For example, these differences could come from different conceptualizations of uncertainty or different estimation methods. Would-be users are left wondering whether the differences in these standard errors imply that some are not valid, and how they should choose between them in practice. As demonstrated by Table 1, these considerations have significant practical implications for applied work. 
We develop a framework that explains how results like those in Table \ref{tab:motivating_differences} can arise, and provides guidance for interpreting and selecting appropriate variance estimators.

\vspace{0.5cm}

\begin{table}[h!]
    \centering
    \footnotesize
    \begin{threeparttable}
        \caption{Comparison of Standard Errors for Synthetic Control Estimator Across Datasets}
        \label{tab:motivating_differences}
        \begin{tabular}{llccccc}
        \toprule
        && & \multicolumn{4}{c}{{Standard Errors}}\\
        Dataset & Treated 
        &  Point & Marginal & Unit & Time & Conditional \\
        &Period& Estimate&& Placebo & Placebo\\
        &  & & (M) & (UP) & (TP) & (C) \\
        \midrule
        \\
        California Smoking& 1989  & -8.459 & 0.412 & 0.245 & 0.083 & 0.048 \\
        \\
        German Unification & 1990  & 0.314 & 0.023 & 0.037 & 0.003 & 0.006 \\
        \\
        Mariel Boatlift & 1980 & 0.069 & 0.009 & 0.006 & 0.008 & 0.006 \\
        \bottomrule
        \end{tabular}
        \begin{tablenotes}
            \small
            \item \textit{Note: This table reports point estimates and standard errors from a truncated dataset ranges up to the first treated time period for each dataset. All subsequent time periods are discarded for all units.} 
        \end{tablenotes}
    \end{threeparttable}
\end{table}

\section{Unified Framework: revisiting variance estimators}

In this section we revisit the previously proposed variance estimators and re-interpret them in a new framework.

\subsection{Set Up}

In this section, we focus on a setting with a single treated unit/period, so that $\sum_{i=1}^N\sum_{t=1}^T W_{it}=1,$ with $(\istar,\tstar)$ denoting the treated unit/period pair, so $W_{\istar,\tstar}=1$ and $W_{i,t}=0$ for $(i,t)\neq (\istar,\tstar)$. For this unit/period pair we observe the treated outcome, $\ystar(1)$, but not the control outcome,
$\ystar(0)$. We have a generic estimator for this control outcome, which we denote by $\hat{Y}_{\istar,\tstar} (0), $ which is a function of control outcomes $Y_{js}(0)$, for all $(j,s)\neq (\istar,\tstar)$. We think of an estimator as a function $\hat{Y}_{it}(0)=g(i,t;\by,\bw)$ that takes in a pair of indices $(i,t)$ and  the matrices $\by$ and $\bw$, and then outputs an estimate $\hat{Y}_{it}(0)$ using only the values in the matrices $\by$ and $\bw$ other than the $(i,t)$ element. The specific estimators that we  use  in this paper to illustrate the ideas are the DID/TWFE estimator, the SC estimator, and the SDID estimator, but the insights also apply to other point estimators such as the augmented synthetic control estimator \citep{ben2021augmented}, the matrix completion estimator \citep{athey2017matrix}, or the triply robust panel estimator \citep{athey2025triply}. For our main cases, we denote the estimators by $g^{\rm TWFE}(\cdot)$, $g^{SC}(\cdot)$, and $g^{\rm SDID}(\cdot)$.
In the Algorithms Appendix--\ref{alg:twfe_estimation_var} for TWFE, \ref{alg:sc_estimation_var} for SC and \ref{alg:sdid_estimation_var} for SDID--we present some details on the functions $g(\cdot)$ for these three cases.

In this section, we describe the three general approaches to variance estimation in the panel data setting that exist in the literature, applied to the case with a single treated unit/period pair. We re-express these in a common framework where all three can be interpreted, at least approximately, as placebo variance estimators where we first estimate residuals and then use averages of squares of the estimated residuals in different placebo exercises to estimate the variance. 

The estimator for the effect on the treated, $\tau=\sum_{i=1}^N\sum_{t=1}^T W_{it}(Y_{it}(1)-Y_{it}(0))$, given a generic estimator $g(i,t;{\bf Y},{\bf W})$ for the missing potential outcome, is
\[\hat\tau\equiv \sum_{i=1}^N\sum_{t=1}^T W_{it} \Bigl(Y_{it}-g(i,t;\by,\bw)\Bigr)\left/
\sum_{i=1}^N\sum_{t=1}^T W_{it} \right.
.\]
The estimated residuals are defined for all pairs $(i,t)$ as
\begin{equation}\label{eq:residuals}
    \hat\varepsilon_{it}\equiv 
Y_{it}-g(i,t;\by,\bw).
\end{equation}

The first approach to variance estimation is related to the classic homoskedastic least-squares regression variance estimator for the TWFE estimator. 
In that case the variance estimator is based on the standard OLS variance for the regression specification
\[ Y_{it}=\alpha_i+\beta_t+\tau W_{it}+\varepsilon_{it}.\]
It can be shown that the homoskedastic OLS variance for this specification is very close to  averaging the squared residuals over all units and time periods other than the treated unit/period pair $(\istar, \tstar)$. This {\it Marginal} (M) variance estimator is $\tsm$ where for all $(i,t)$
\[\hat\sigma^{2,M}_{it}\equiv\frac{1}{NT-1} \sum_{s=1}^{T}\sum_{j=1}^N \mathbb{1}_{(j,s)\neq (i,t)} \hat\varepsilon_{js}^2\hskip2cm
\mathbf{(Marginal,\: M).}\]
The advantage of expressing this estimator in terms of averaged squared residuals is that it extends naturally to other estimators, such as SC, SDID, and beyond.

The second variance estimator averages the squared residuals from all units in the treated period $\tstar$, excluding the treated unit $\istar$, implying a unit-exchangeability assumption. The  {\it Unit Placebo} (UP) variance estimator for $\sigma^2_{it}$ is
\[\hat \sigma^{2,UP}_{it}\equiv\frac{1}{N-1} \sum_{j=1}^{N}  \mathbb{1}_{j\neq i } \hat\varepsilon_{jt}^2\hskip2cm
\mathbf{(Unit \; Placebo,\: UP).}\]
Methods closely related to this approach, which sometimes focus on testing rather than variance estimation, have been used in the synthetic control literature \citep{abadie2010synthetic, doudchenko2016balancing}. In fact, \citet{abadie2010synthetic} used a unit placebo-style estimator to conduct inference in their seminal synthetic control paper.

The third approach to variance estimation is based on exploiting the time-exchangeability rather than unit-exchangeability, and averages the squared residuals for treated unit $\istar$ over all time periods other than treated period $\tstar$. This {\it Time Placebo} (TP) variance estimator for $\sigma^2_{it}$  is
\[\hat \sigma^{2,TP}_{it}=\frac{1}{T-1} \sum_{s=1}^{T} \mathbb {1}_{s\neq t} \hat \varepsilon_{is}^2\hskip2cm
\mathbf{(Time \; Placebo,\: TP).}\]
Methods based on this approach have also been used extensively in the more recent synthetic control literature \citep{doudchenko2016balancing, chernozhukov2021exact}, sometimes with connections to the conformal inference literature \citep{lei2021conformal, chernozhukov2021exact} .

\subsection{Motivational Simulation Study}

In the introduction, we showed some of the variation in the different standard errors for three real data sets.
Here, we examine the repeated-sampling performance of these variance  estimators in a realistic simulation setting, based on simulation designs from \citet{arkhangelsky2021synthetic}. As a baseline, we use the same three data sets for which we reported results in Table \ref{tab:motivating_differences}, the California smoking data, the German Unification data, and the Mariel Boatlift data. We remove the  periods where one unit is treated so we only have control outcomes.  In addition we use four panel data sets for which there are no true treatments. Of these, three are derived from the CPS, with $N=50$ and $T=40$, typical of the data sets used in this literature. 
These baseline panels consist of yearly state-level data from all $N=50$ states on log wages, the state unemployment rate, and average weekly hours of work for the $T=40$ years from 1979-2019, as provided by \citet{arkhangelsky2021synthetic}'s synthdid package.
In addition we use the data on per capita GDP from \citet{feenstra2015Penn}'s Penn World Table database, which contains information on relative levels of income, output, inputs, and productivity for 167 countries between 1950 and 2011. We restrict this dataset to 111 countries for which we have 48 years of data (between 1959 and 2007). The outcome matrix, $\by$, contains values of log annual real GDP.\footnote{Note that for computational reasons, in the variance comparison exercise in Tables \ref{table:variance_comparison_table_sc_new_format} and \ref{table:coverage_comparison_table_sc} we restrict the simulation to a matrix of 50 by 40.} 

The specific simulation procedure we use is described in  Algorithm \ref{alg:placebo_studies}. We start by generating realistic simulated panels that share important characteristics to our real baseline panels, following a slightly modified version of the simulation design in \citet{arkhangelsky2021synthetic}, as described in Appendix Algorithm \ref{alg:dgp_estimation}. 
We then randomly select a cell  $(i,t)$ to serve as the pseudo-treated unit and period, and compute the M, UP, TP and the to-be-defined C standard errors. 

To summarize the results, we first calculate the standard deviation of the estimated $\hat{\tau}_{\istar\tstar}$ over all $R$ pairs (labeled \textit{S.D.($\hat{\tau}_{\istar\tstar})$} in Table \ref{table:variance_comparison_table_sc_new_format}). Next, we calculate the mean and standard deviation of the standard errors over all  $R$ pairs. If the estimator is performing well on average, then the mean of the standard errors should be approximately equal to the empirical standard deviation of $\hat{\tau}_{\istar\tstar}$. We present results for the SC estimator in Table \ref{table:variance_comparison_table_sc_new_format}, and present results for the TWFE and SDID estimators (Appendix \ref{table:variance_comparison_table_twfe_new_format} for TWFE and \ref{table:variance_comparison_table_sdid_new_format} for SDID). 

\vspace{0.5cm}

\begin{algorithm}[H] 
\SetAlgoLined
\KwData{Panel data $\by$, Treatment matrix $\bw$, Estimator $\in$ (TWFE, SC, SDID), Number of simulations $R$}
\KwResult{$R$ estimates of $\hat{\tau}_{it}, \sqrt{\tsmit}$, $\sqrt{\tsuit}$, $\sqrt{\tstit}$, $\sqrt{\tscit}$}

Estimate baseline DGP parameters from real panel data $\by$ using the procedure described in Appendix Algorithm \ref{alg:dgp_estimation} \; 

\For{simulation $r = 1$ to $R$}{
    Generate simulated panel $\by_r$ ($N \times T$) using estimated baseline parameters\;
    Normalize $\by_r$ to mean 0, standard deviation 1\;
    Create treatment matrix $\bw_r$ with single treated unit $(\istar, \tstar)$ chosen randomly, such that $W_{\istar\tstar} = 1$ and all other cells = 0\; 
    Compute $\widehat{Y}_{\istar \tstar}(0)$ using \textit{Algorithm `Estimator'} (see appendix)\;
    Calculate $\hat{\tau}_{\istar\tstar} = \ystar - \widehat{Y}_{\istar\tstar}(0)$, save estimate \;
    Calculate four standard errors for $\hat{\tau}_{\istar\tstar}$: $\sqrt{\tsmit}$, $\sqrt{\tsuit}$, $\sqrt{\tstit}$, $\sqrt{\tscit}$, save estimates\;
}

\caption{Variance Estimation Placebo Exercise}\label{alg:placebo_studies}
\end{algorithm}

\begin{table}[H]
\renewcommand\thetable{2.a}
\renewcommand{\thetable}{2.a}
\centering
\caption{Standard Errors for the Synthetic Control Estimator}
\begin{tabular}{lccccc}
\hline
&& \multicolumn{4}{c}{Average Standard Errors}
\\
&& \multicolumn{4}{c}{(standard deviation)}
\\
Data Set & S.D.($\hat{\tau}_{\istar\tstar}$)  & \multicolumn{1}{c}{UP}
& \multicolumn{1}{c}{TP}
& \multicolumn{1}{c}{M}
& \multicolumn{1}{c}{C}\\
\hline
California Smoking & 0.69 & 0.65 & 0.60 & 0.70 & 0.62 \\
& & (0.25) & (0.36) & (0.04) & (0.50) \\
Germany & 0.98 & 0.84 & 0.81 & 1.03 & 0.74 \\
& & (0.62) & (0.62) & (0.06) & (0.87) \\
Mariel Boat & 1.13 & 1.03 & 0.88 & 1.17 & 0.90 \\
& & (0.53) & (0.77) & (0.12) & (0.85) \\
CPS hours & 1.11 & 0.97 & 0.96 & 1.05 & 0.93 \\
& & (0.40) & (0.45) & (0.02) & (0.61) \\
CPS log wage & 0.93 & 0.86 & 0.85 & 0.92 & 0.84 \\
& & (0.34) & (0.31) & (0.02) & (0.48) \\
CPS urate & 0.71 & 0.71 & 0.70 & 0.74 & 0.69 \\
& & (0.23) & (0.25) & (0.02) & (0.34) \\
PENN & 0.52 & 0.53 & 0.48 & 0.55 & 0.50 \\
& & (0.14) & (0.26) & (0.02) & (0.32) \\
\hline
\end{tabular}
\label{table:variance_comparison_table_sc_new_format}
\begin{tablenotes}
\small
\item \textit{Note: This table reports  standard errors based on the four variance estimators, with their standard deviations in parentheses. The column S.D.($\hat{\tau}$) reports the standard deviation of the point estimate over the repeated samples.} 
\end{tablenotes}
\end{table}

Two key patterns emerge in Table \ref{table:variance_comparison_table_sc_new_format} that hold across these data sets and estimators. The first finding is that all four standard errors are approximately right on average, across all estimators (see Appendix \ref{table:variance_comparison_table_twfe_new_format} for TWFE and \ref{table:variance_comparison_table_sdid_new_format} for SDID) and all data sets.

Consider the California smoking data. Across all simulations, the standard deviation of the estimator is $0.69$. 
The average of the M standard errors is $0.70$, close to the actual standard deviation. The same holds for the other three standard errors: $0.65$ for the average of the UP standard error, $0.60$ for the average of the TP standard error, and $0.62$ for the average of the C standard error. 

The second finding is that although the standard errors are right on average, they differ substantially in any given case. As a result, the choice between the variance estimators matters. In particular, across simulations, the standard deviation of the standard errors is quite different for the four standard errors, M, UP, TP, and C. For example, for the California Smoking data, the standard deviation is $0.04$ for the M-based standard errors, $0.25$ for UP,  $0.36$ for TP, and $0.50$ for C. 
In addition, the correlations between the variance estimators are quite low and often negative.

\begin{table}[H]
\renewcommand\thetable{2.b}
\centering
\caption{Coverage Rates for Nominally 95\% Confidence Intervals for  the Synthetic Control Estimator}
\begin{tabular}{p{6cm}cccc}
\hline
Data Set & \multicolumn{1}{c}{UP}
& \multicolumn{1}{c}{TP}
& \multicolumn{1}{c}{M}
& \multicolumn{1}{c}{C}\\
\hline
California Smoking & 0.93 & 0.92 & 0.94 & 0.92 \\
Germany & 0.91 & 0.93 & 0.95 & 0.92 \\
Mariel Boat & 0.94 & 0.87 & 0.94 & 0.88 \\
CPS hours & 0.92 & 0.92 & 0.93 & 0.94 \\
CPS log wage & 0.96 & 0.92 & 0.95 & 0.95 \\
CPS urate & 0.95 & 0.94 & 0.95 & 0.92 \\
PENN & 0.92 & 0.94 & 0.93 & 0.95 \\
\hline
\end{tabular}
\label{table:coverage_comparison_table_sc}
\end{table}

We also report the coverage for confidence intervals based on each standard error in Table \ref{table:coverage_comparison_table_sc}, calculated as the fraction of pairs such that the difference between the imputed value $\widehat{Y_{it}}(0)$ and the actual value $Y_{it}(0)$ is less than 1.96 times the standard error in absolute value. Note that the validity of the confidence intervals relies on strong distributional assumptions, even in large samples, because there is only a single treated unit/period pair. Nevertheless, the coverage rates are quite good.  For example, the coverage rates for the California smoking data are $0.94$ based on the M standard errors, $0.93$ based on the UP standard errors,  $0.92$ based on the TP standard errors, and $0.92$ for the C standard errors. Similar results are obtained for the other six data sets, and also hold for the other two estimators, TWFE and SDID (see Appendix \ref{table:coverage_comparison_table_twfe} for TWFE and \ref{table:coverage_comparison_table_sdid} for SDID). 

\section{A New Variance Estimator: Conditional Variance}\label{section:new estimator}

In this section, we discuss the statistical problem of estimating the variance of the untreated potential outcome for a single treated unit-time pair. We begin with a simplified setting to isolate core conceptual challenges and then extend to a more realistic residual structure that allows for unit and time heterogeneity. Our goal is to determine when and why different standard errors--Marginal (M), Unit-Placebo (UP), Time-Placebo (TP), or Conditional (C)--are appropriate.
Initially, we focus on a very stylized version of the variance estimation problem that highlights the core conceptual issues that are our primary focus. 

\subsection{Formalization}

First, we decompose the control potential outcome as $Y_{it}(0)=\mu_{it}+\varepsilon_{it}$, where $\mu_{it}\equiv\mme[Y_{it}(0)]$ is the systematic component for unit $i$ in period $t$ and $\varepsilon_{it}\equiv Y_{it}(0)-\mu_{it}$ is the error or idiosyncratic component with marginal expectation equal to zero.
The decomposition separates the predictable structure ($\mu_{it}$) from the unpredictable noise ($\varepsilon_{it}$). In large samples, which can be a combination of large $N$ and large $T$, a good estimator will estimate $\mu_{it}$ well, so $\hat\mu_{it}-\mu_{it}=o_p(1)$.
Accordingly, we ignore the systematic component and assume $\mu_{it} = 0$ for all $i,t$.
In that case a natural estimator for the missing outcome is $\hat{Y}_{\istar\tstar}(0)=0$, and the resulting $O_p(1)$ estimation error is \[ \ystar(0)-\hat{Y}_{\istar\tstar}(0)=\varepsilon_{\istar\tstar}.\]
More generally, the error would also have a component $\hat\mu_{it}-\mu_{it}$, which will usually be smaller than the $\varepsilon_{\istar\tstar}$ component.

The challenge is to estimate the variance of $\varepsilon_{\istar\tstar}$  on the basis of the $N\times T-1$ values of the idiosyncratic components $\varepsilon_{it}$ for $(i,t)\neq (\istar, \tstar)$.

The first assumption we make, and which we maintain throughout the paper, is that the unobserved components are independent of the treatment assignment 
$\bw$.
\begin{assumption}\label{assumption:ind}{\sc Independence}
\[ \mathbf{\varepsilon}\ \indep\ \bw.\]
\end{assumption}

Next, we assume exchangeability of the error terms $\varepsilon_{it}$. See \citet{de2017theory} for a general discussion on exchangeability, and \citet{aldous1981representations, lynch1984canonical, 
mccullagh2000resampling} for definitions in settings with arrays.

\begin{assumption}\label{assumption:rce}{\sc Row and Column  Exchangeability}\\
The $N\times T$ matrix $\mathbf{\varepsilon}$ is row and column and exchangeable (RCE).
\end{assumption}
This assumption combines two separate assumptions: {column or time exchangeability} (CE), which arises when the columns of the matrix $\varepsilon$, corresponding to time periods are exchangeable, and  {row  or unit exchangeability} (RE) which arises when the rows of $\varepsilon$, corresponding to the units, are exchangeable.
In general, unit exchangeability is commonly assumed. Time exchangeability for the raw outcome data is typically not plausible, although it may be a more reasonable approximation for the residuals as opposed to the raw outcomes. Moreover, it can be weakened  by allowing for some autocorrelation.

The implication of the $N\times T$ matrix $\varepsilon$ being both row and column  exchangeable  is that there is a random  $N\times T$  matrix $\varepsilon^*$ with typical element
\[ \varepsilon^*_{it}=f(\kappa,\nu_i,\xi_t,\eta_{it}),\]
with all $(\kappa,\nu_i,\xi_t,\eta_{it})$ jointly independent, such that the distributions of $\varepsilon$ and $\varepsilon^*$ are identical \citep{aldous1981representations, lynch1984canonical, 
mccullagh2000resampling}. In this representation \( \kappa \) is a global latent variable that is common across all observations in the matrix. For ease of exposition we drop this conditioning going forward.  The variable \( \nu_i \) captures unit-specific heterogeneity, while \( \xi_t \) accounts for time-specific heterogeneity. Finally, \( \eta_{it} \) is an idiosyncratic shock that varies across both units and time periods. Together, these components generate the matrix of residuals.

Next we define the mean and variance for four conditioning sets ($\emptyset$ (implicitly conditioning on $\kappa$),
 $\{\nu_i\}$,
  $\{\xi_t\}$
and $\{\nu_i,\xi_t\}$),
the conditional expectations and variances of $\varepsilon^*_{it}$:
\[ \mu\equiv \mathbb{E}[\varepsilon^*_{it}],\qquad
\sigma^{2}\equiv \mathbb{E}[(\varepsilon^*_{it}-\mu)^2] \quad \textit{(marginal)} \]
\[ \mu_\nu(\nu_i)\equiv \mathbb{E}[\varepsilon^*_{it}|\nu_i],\qquad
\sigma^{2}_\nu(\nu_i)\equiv \mathbb{E}[(\varepsilon^*_{it}-\mu_\nu(\nu_i))^2|\nu_i] \quad \textit{(unit-conditional)}\]
\[ \mu_\xi(\xi_t)\equiv \mathbb{E}[\varepsilon^*_{it}|\xi_t],\qquad
\sigma^{2}_\xi(\xi_t)\equiv \mathbb{E}[(\varepsilon^*_{it}-\mu_\xi(\xi_t))^2|\xi_t] \quad \textit{(time-conditional)}\]
\[ \mu(\nu_i,\xi_t)\equiv \mathbb{E}[\varepsilon^*_{it}|],\qquad
\sigma^{2}(\nu_i,\xi_t)\equiv \mathbb{E}[(\varepsilon^*_{it}-\mu(\nu_i,\xi_t))^2|\nu_i,\xi_t] \quad \textit{(fully conditional)}\]

\begin{assumption}{\sc (Independence)}\label{assumption:exp}
For all $\nu,\xi$
 \[
\mu(\nu,\xi)=0,\]
\end{assumption}
(The subscripts on $\mu_\nu(\cdot)$ and $\mu_\xi(\cdot)$, and between $\sigma^2_\xi(\cdot)$ and $\sigma^2_\nu(\cdot)$, for for notational precision, to distinguish between the conditioning sets.)

The following result establishes that under independence and exchangeability, all four  variances (marginal, unit-conditional, time-conditional, and fully conditional) have the same expectation.

\begin{proposition}\label{prop_var}{\sc Variances}\\ Suppose that Assumptions \ref{assumption:ind}, \ref{assumption:rce} and 
\ref{assumption:exp}
hold. Then\\
$(i)$
for all $\mu,\xi$,
\[\mu=\mu_\nu(\nu)=\mu_\xi(\xi)=0,\]
$(ii)$
\[\mmv\left(\varepsilon_{it}^2\right)= \sigma^{2}=\mme\left[\sigma^{2}_\nu(\nu_i)\right]=\mme\left[\sigma^{2}_\xi(\xi_t)\right]=  \mme\left[\sigma^{2}(\nu_i,\xi_t)\right],\]
$(iii)$
with the partial ranking
\[\mmv(\sigma^{2})\quad \leq\quad \mmv(\sigma^{2}_\nu(\nu_i)),\mmv(\sigma^{2}_\xi(\xi_t))\quad \leq\quad \mmv(\sigma^{2}(\nu_i,\xi_t)). \]
The ranking between 
$\mmv(\sigma^{2}_\nu(\nu_i))$ and $\mmv(\sigma^{2}_\xi(\xi_t))$ is not generally determined.
\label{prop1}
\end{proposition}

See proof in Appendix \ref{appx:proofs}. The implication is that under the assumption that the $\varepsilon_{it}$ have (conditional) expectation zero,  $(i)$ all four variances are valid in the sense of having expectation equal to the second moment of $\varepsilon_{it}$,  and $(ii)$  the values will generally  be different in any particular sample.
Given that all four variances are valid in this case, the question arises which variance researchers should use. 
In practice, choosing a variance estimator is even more challenging than choosing a variance, because the variances depend on unknown quantities that must themselves be estimated. To separate some of these issues, we'll first address the choice between these four oracle variances.

This question is somewhat similar to the famous ancillarity example from David Cox \citep{cox1958some, cox1971choice} which we summarize here briefly. Suppose we are interested in measuring the length of a physical object which has true length $\theta$. The Cox thought experiment consists of two stages. In the first stage, the experimenter flips a coin. If it comes up heads, the experimenter will use an accurate measuring instrument that will lead to a measure that has a normal distribution ${\cal N}(\theta,0.01)$. If the coin comes up tails, the experimenter will use an inaccurate measuring instrument that will lead to a measure that has a normal distribution ${\cal N}(\theta,100)$. 
Given a measure $X$, and given that the coin came up $Y$ (either $Y=H$ or $Y=T$), what measure of uncertainty should the experimenter report? One option is the marginal variance, $V^M=1/2(0.01+100)=50.005$. Another option is the conditional variance, $V^C(Y)$, equal to $0.01$ if $Y=H$ and $100$ if $Y=T$. 
Both are valid in the sense that $\mathbb{E}[V^C(Y)]=V^M=\mathbb{E}[(X-\theta)^2].$
Cox argues that because of ancillarity of the choice of instrument, the experimenter should always report the conditional variance $V(Y)$ rather than the marginal variance $V^M=\mathbb{E}[V^C(Y)]$. 
Where does this show benefits in practice? Suppose one wants to test whether the length of the object is 10 at the 5\% level, and in fact the length of the object is 10.5. Using the conditional variance would lead to a power of 0.5 (essentially 1 whenever the precise measuring instrument is used, but almost 0 when the inaccurate instrument is used). The power based on the marginal variance would be approximately 0.10, about twice the size of the test.

The three previously introduced, and commonly used, variance estimators, $\hat\sigma^{2,\marg}_{it}$, 
$\hat\sigma^{2,\unit}_{it}$,  and $\hat\sigma^{2,\ttime}_{it}$,  estimate three of these (conditional) variances. This is summarized in the next proposition.
\begin{proposition}
 Suppose that 
  Assumptions \ref{assumption:ind}, \ref{assumption:rce} and 
\ref{assumption:exp} hold, and that $\hat{Y}_{it}=0$.
Then
\[ \hat \sigma^{2,\marg}_{it}-\sigma^2=o_p(1),\qquad  \hat \sigma^{2,\unit}_{it}-\sigma^2_\xi(\xi_t)=o_p(1),
\qquad {\rm and}\quad
\hat \sigma^{2,\ttime}_{it}-\sigma^2_\nu(\nu_i)=o_p(1).\]
\label{prop2}
\end{proposition}

See proof in Appendix \ref{appx:proofs}.
Let us revisit 
Table \ref{table:variance_comparison_table_sc_new_format} in the light of these correspondences.
Indeed, as shown in Table \ref{table:variance_comparison_table_sc_new_format}, the three previously proposed variance estimators, M, UP, and TP, estimate the same marginal variance on average. We also observe that the UP and TP standard errors have a larger variance than the M standard errors. Furthermore, we note that the ranking of the UP and TP standard errors in terms of variance depends on the particular data-generating process.

The problem is that these three previously proposed variance estimators correspond to inferior variances by the ancillarity principle, and that this principle does not lead to a clear ranking of these three variance estimators. 
A second challenge is that there is no natural estimator for the superior oracle variance $\sigma^2(\kappa,\nu_i,\xi_t)$ because we cannot average squared residuals for the treated unit/period because there is only a single such unit/period pair.

\subsection{A New Variance Estimator}

The UP and TP variance estimators allow for general forms of heteroskedasticity over time and across units, respectively, but not both. In practice, both types of heteroskedasticity may be present. 
However, we cannot condition both on time and unit, because we only have a single observation for each unit/time pair. 
Our new conditional (C) variance estimator therefore, relies on some assumptions that put structure on the nature of the heteroskedasticity. The specific structure that we use is in the form of a two-way-fixed-effect model:
\begin{assumption}
\begin{equation}\label{varmodel}\sigma^{2}_{it}=\exp(\kappa+\nu_i+\xi_t).\end{equation}
\end{assumption}
The $\nu_i$ captures the heteroskedasticity across units, and the $\xi_t$ captures the heteroskedasticity over time.

To estimate the variance we first estimate a linear regression of the logarithm of the square of the residuals
$\hat\varepsilon_{it}^2$ on unit and time fixed effects:
\[ \ln\left(\hat\varepsilon^2_{it}\right)=\gamma+\nu_i+\xi_t+\eta_{it}.\]
Here we normalize the $\nu_i$ and $\xi_t$ to be mean zero.
We then estimate the conditional variance for the  unit/period pair $(i,t)$ as\footnote{The adjustment factor $\exp(1.2704)$ in this variance estimator adjusts for the fact that if $\varepsilon$ has a standard Normal distribution ${\cal N}(0,1)$, then
$\mathbb{E}\left[\ln\left(\varepsilon^2\right) \right]=\int_{-\infty}^{\infty} \ln(z^2)\frac{1}{\sqrt{2\pi}}\exp\left(-z^2/2\right)dz=\frac{\Gamma(1/2)}{\sqrt{\pi}}\left(\ln(2)+\psi(1/2) \right)\approx \exp(-1.2704).$ 
This uses the change of variables result which implies that
\[ \int_0^{\infty}
\ln(y^2)\exp(-y^2/2)dy
=\frac{1}{\sqrt{2}}\int_0^{\infty}\ln(2t) t^{-1/2} \exp(-t)dt.\]
} 
\[ \hat{\sigma}^{2,C}_{it}=\exp\left(1.2704+\hat\kappa+\hat\nu_i+\hat\xi_t\right)\hskip2cm
\mathrm{\bf (Conditional, C).}\]

\paragraph{Comparing with other estimators} We can now revisit Table \ref{table:variance_comparison_table_sc_new_format} and observe that the new estimator $\tsc$ is $(i)$ on average as good as other estimators, and $(ii)$ has a larger standard deviations than other estimators. This aligns with 
Proposition \ref{prop2}, on average, and under the exchangeability assumption, the conditional estimator is as good as the other estimators, but the variance of the standard errors is larger.

We note that although the model for the variance in Equation (\ref{varmodel}) allows for both unit and time heteroskedasticity, it is a relatively simple one with only additive unit and time effects. One could easily generalize this model to allow for a more flexible factor structure,
\[ \sigma_{it}^2=\exp\left(\sum_{r=1}^R \alpha_{ir}\beta_{tr}\right).\]
Especially in settings with a large number of units and time periods it may be useful to consider such  models. We defer the discussion of such models for the variance to future work.

\section{Evaluation through Simulations}

In this section, we assess the properties of the new C (Conditional) variance estimator relative to the UP, TP and M variance estimators in finite samples using a simulation study. As in the motivational example above, we consider a setting with a single treated unit/period pair; then, we display the power curve for estimating an imposed treatment $\tau \in [-3;3]$. 
We present two sets of simulations.
In the first, illustrative set of simulations, we consider a very stylized data-generating process that illustrates the points about power curves we wish to make. In the second set of simulations, we match the design to some of the data sets we have used earlier to illustrate our insights to show that these issues are relevant in realistic settings.

We consider a stylized data-generating process that matches the model for the variance in Equation (\ref{varmodel}),  with design parameters $K_\xi$ and $K_\nu$,
\begin{equation} \label{eq:twfe_resid}
    Y_{it}(0)
    =\varepsilon_{it}=
 \eta_{it}\exp(\nu_i/2+\xi_t/2),
\end{equation}
 with $\nu_i$, $\xi_t$, and $\eta_{it}$ independent,
 and  $\eta_{it}\sim{\cal N}(0,1)$
 so that  
\[ \sigma_{it}^2=\mmv(\varepsilon_{it}|\nu_i,\xi_t)=\exp(\nu_i+\xi_t),\]
as in Equation (\ref{varmodel}).
We choose rescaled Chi-squared distributions for $\exp(\nu_i)$ and $\exp(\xi_t)$:
\[\exp(\nu_i)\sim{\cal X}^2(K_\nu)/K_\nu
,\quad
 \exp(\xi_t)\sim{\cal X}^2(K_\xi)/K_\xi
 ,\]
 so that $\mme[\exp(\nu_i)]=1,$ $\mmv(\exp(\nu_i))=2/K_\nu$,
   $\mme[\exp(\xi_t)]=1$, and $\mmv(\exp(\xi_t))=2/K_\xi$.
 The marginal variance of $\varepsilon_{it}$ is $\sigma^2=\mme[\sigma^2_{it}]=1$ for all unit/period pairs $(i,t)$.
The three conditional variances are
\[   \sigma^{2}_\nu(\nu_i)=\exp(\nu_i),\quad\sigma^{2}_\xi(\xi_t)= \exp(\xi_t) ,\quad \sigma^{2}(\nu_i,\xi_t)=\exp(\nu_i+\xi_t)
.\]
All four of these  variances (the marginal and the three conditional ones) have expectation equal to the marginal variance of $\varepsilon_{it}$, $\mmv(\varepsilon_{it}^2)=1$, but they have quite different variances:
\[ \mmv(\sigma^{2})=0,\quad \mmv(\sigma^{2}_\nu(\nu_i))=\mmv(\exp(\nu_i))=2/K_\nu,\quad\mmv(\sigma^{2}_\xi(\xi_t))=\mmv(\exp(\xi_t))=2/K_\xi,\]and
\[\mmv(\sigma^{2}(\nu_t,\xi_t))=(1+\mmv(\exp(\nu_i)))(1+\mmv(\exp(\xi_t)))-1=2/K_\nu+2/K_\xi+4/(K_\nu K_\xi).\]

The key choice of the degrees of freedom parameters $K_\xi$ and $K_\nu$ allows us to systematically control the extent of heteroskedasticity in both dimensions. When $K_\xi$ or $K_\nu$ approach infinity (corresponding to $\mmv(\xi_t)$ or $\mmv(\nu_i)$ approaching zero), the corresponding rescaled chi-squared distributions become degenerate at 1, yielding homoskedastic variances in that dimension. But when these parameters approach 1, the variance of the chi-squared components increases dramatically, creating substantial heteroskedasticity. Given the direct mapping between degree of freedom and variance of the chi-squared distribution, we mainly focus on variances, defined as $\mmv(\sigma^{2}(\cdot))$ for ease of understanding in the rest of this section. 

This framework enables us to isolate the performance of each variance estimator under different patterns of heteroskedasticity: unit-only heteroskedasticity (high $\mmv(\nu_i)$, zero $\mmv(\xi_t)$), time-only heteroskedasticity (zero $\mmv(\exp(\nu_i))$, high $\mmv(\exp(\xi_t))$), or heteroskedasticity in both dimensions (high $\mmv(\nu_i)$ and high $\mmv(\xi_t)$). The conditional variance $\sigma^{2}(\kappa,\nu_i,\xi_t)=\exp(\nu_i+\xi_t)$ captures the full interaction between unit and time heterogeneity, making it the most variable but also the most informative for the specific treated cell.

\subsection{Illustrative Simulations}\label{section:baseline}

We consider four simulation designs using this data generating process. As described, the distribution for $Y_{it}$ is Gaussian ${\cal N}(0,\sigma_{it}^2)$, with potentially heterogeneous variances $\sigma^2_{it}=\exp(\nu_i+\xi_t)$.
We compare four cases that directly illustrate the theoretical predictions about variance estimator performance. We use a variance of 2 for $\exp(\mu_i)$ or $\exp(\xi_t)$  (or equivalently, degrees of freedom equal to 1) to represent substantial heteroskedasticity in our ``high'' $\mmv$ condition, and a variance of 0 to capture the homoskedastic case.

\textbf{Case 1: Homoskedastic baseline ($\mmv(\exp(\nu_i))=\mmv(\exp(\xi_t))=0$).} Under homoskedasticity, all variance estimators should perform similarly since there is no heterogeneity to exploit or account for.

\textbf{Case 2: Unit heteroskedasticity only ($\mmv(\exp(\nu_i))=2$, $\mmv(\exp(\xi_t))=0$).} With heterogeneity limited to the unit dimension, we expect $\hst$ to exhibit high variability across simulations as it attempts to capture unit-specific variance patterns. Meanwhile, $\hsu$ and $\hsm$ should remain stable since they either avoid or average over this source of variation.
In this case the C and TP standard errors should be preferred.

\textbf{Case 3: Time heteroskedasticity only ($\mmv(\exp(\nu_i))=0$, $\mmv(\exp(\xi_t))=2$).} Conversely, when heterogeneity is purely temporal, $\hsu$ becomes highly variable as it captures time-specific variance patterns, while $\hst$ and $\hsm$ maintain stability. Now the C and UP standard errors should be preferred.

\textbf{Case 4: Two-way heteroskedasticity ($\mmv(\exp(\nu_i))=\mmv(\exp(\xi_t))=2$).} This most challenging scenario features heterogeneity in both dimensions. Here, both $\hst$ and $\hsu$ become variable, but $\hsc$ should exhibit the highest variability as it attempts to capture the full interaction $\nu_N\xi_T$ specific to the treated cell. Only $\hsm$ maintains stability through its averaging across all cells. Here the C standard errors should be preferred for power reasons.

These four cases provide a systematic test of how each estimator responds to different heteroskedasticity patterns, with power curves revealing which estimator best balances precision and appropriate uncertainty quantification under each scenario.

\begin{figure}[htbp!]
    \centering
    \caption{Power Curves: Illustrative Simulations Using SC Estimator.}        
     \begin{subfigure}{0.48\textwidth}
        \centering
        \includegraphics[width=\textwidth]{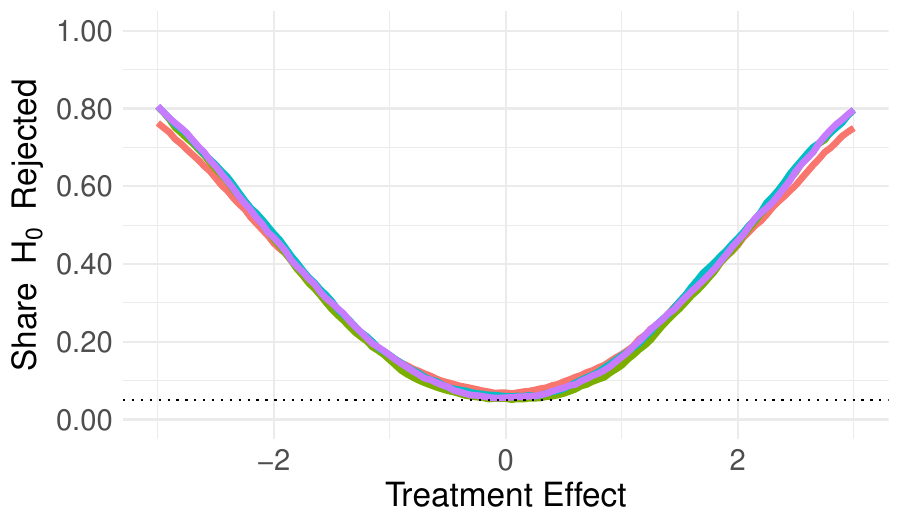}
        \caption{$\mmv(\exp(\nu_i))=\mmv(\exp(\xi_t))=0$}
    \end{subfigure}
\begin{subfigure}{0.48\textwidth}
        \centering
        \includegraphics[width=\textwidth]{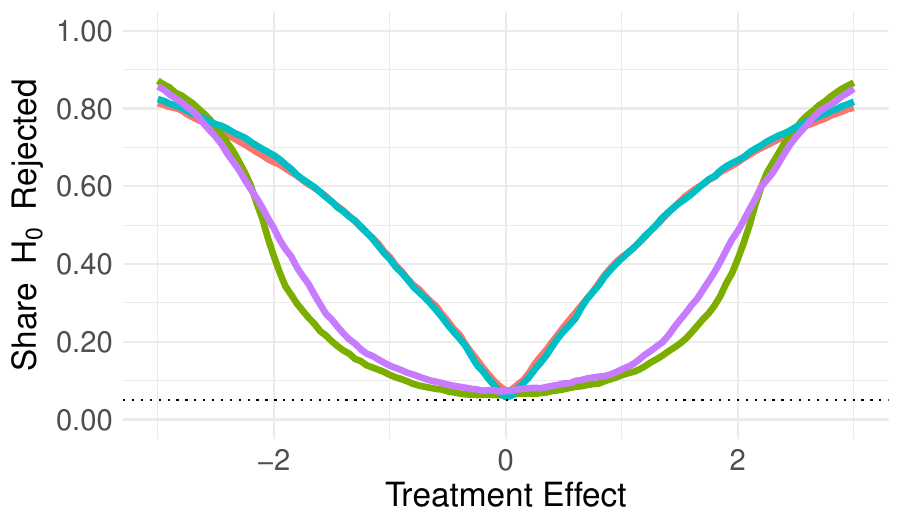}
        \caption{$\mmv(\exp(\nu_i))=2$, $\mmv(\exp(\xi_t))=0$}
    \end{subfigure}

 \begin{subfigure}{0.48\textwidth}
        \centering
        \includegraphics[width=\textwidth]{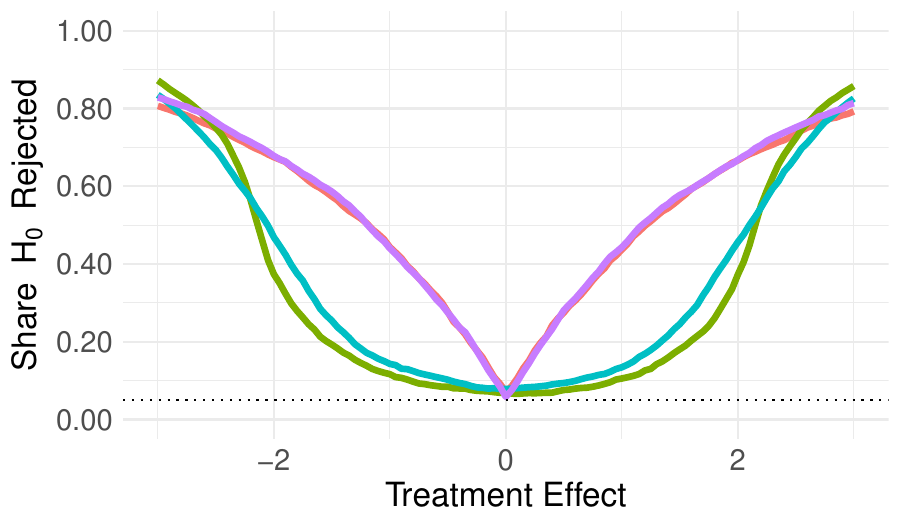}
        \caption{$\mmv(\exp(\nu_i))=0$, $\mmv(\exp(\xi_t))=2$}
    \end{subfigure}
     \begin{subfigure}{0.48\textwidth}
        \centering
        \includegraphics[width=1.2\textwidth]{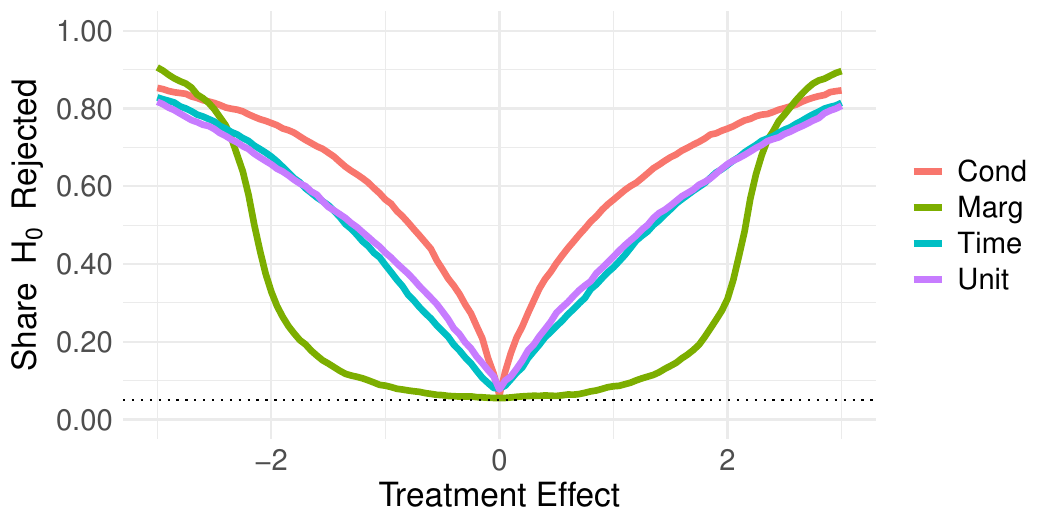}
        \caption{$\mmv(\exp(\nu_i))=\mmv(\exp(\xi_t))=2$}
    \end{subfigure}
    \label{fig:simple_sim_studies_sc}
\end{figure}

Figure \ref{fig:simple_sim_studies_sc} presents power curves computed using data from each of the four simulation designs. As predicted, the power curves show that under homoskedasticity (Panel A), all variance estimators perform similarly. If there is heterogeneity by units but not by time periods, as in Panel B, the power of the TP and C variance estimators is superior. If there is heterogeneity by time but not by units, as in Panel C, the UP and C variance estimators have superior power. Finally, if there is heterogeneity of both types, our new C variance estimator has superior power that to  all three previously proposed variance estimators. Appendix Figures \ref{fig:simple_sim_studies_sdid} and \ref{fig:simple_sim_studies_twfe} present similar results for the TWFE and SDID estimators.

\subsection{Semi-synthetic Data Simulations}

Now that we have demonstrated that the estimators' performance differ when the sources of heteroskedasticity differ, we return to the seven real datasets we previously considered to demonstrate that these performance differences persist under realistic levels of heteroskedasticity.

First, we wish to see whether the residuals in these data sets exhibit the type of heteroskedasticity that suggest that either the UP or TP standard errors might be superior to the M standard errors or to each other, or that the C standard error could be superior to both UP and TP standard errors. 

To do so, we revisit Equation \ref{eq:twfe_resid}. We estimate the $\nu_i$ and $\xi_t$ 
by fitting a two-way-fixed-effect regression to the logarithm of $\hat\varepsilon_{it}^2$, rescaling the implied estimates $\exp(\hat\nu_i)$ and $\exp(\hat\xi_t)$ so they have mean one, and then calculating their variances. Details are provided in Section \ref{appendix:var} in the appendix. We report those estimated variances $\hat{\mmv}(\exp(\hat\nu_i))$ and $\hat{\mmv}(\exp(\hat\xi_t))$ in Table \ref{tab:df_chisq_transformed}.
In Table \ref{tab:df_chisq_transformed}, the top three rows use the same datasets as in Table \ref{tab:motivating_differences}, where one unit is treated in the last period. These three datasets  exhibit high unit heteroskedasticity ($\hat{\mmv}(\exp(\hat\nu_i))$ greater than one). The PENN dataset shares this feature, too. The German Unification data additionally have quite substantial time heteroskedasticity. In contrast, the CPS data exhibit rather different characteristics, with both small $\hat{\mmv}(\exp(\hat\nu_i))$ and $\hat{\mmv}(\exp(\hat\xi_t))$ indicating little heteroskedasticity in both dimensions.

\begin{table}[h!]
\centering
\renewcommand\thetable{3}
\caption{Variance of $\nu$ and $\xi$ using standardized $\varepsilon_{it}$} 
\label{tab:df_chisq_transformed}
\begin{tabular}{llcc}
  \hline
 && \multicolumn{2}{c}{Heteroskedasticity}\\
 && Unit & Time\\
 $[N,T]$& Dataset & $\hat{\mmv}(\exp(\hat\nu_i)) $ & $\hat{\mmv}(\exp(\hat\xi_t)) $  \\
  \hline
$[39,20]$ & California Smoking & {4.64} &  {0.36} \\ 
$[17,31]$ & German Unification  &  {2.14} &  {1.30} \\ 
$[44,8]$ & Mariel Boatlift  & 1.26 &  {0.19} \\ 
\hline
 $[50,40]$ & CPS Hours & 0.47 & 0.15 \\ 
 $[50,40]$ & CPS Log Wage & 0.39 & 0.26 \\ 
 $[50,40]$ & CPS Unemployment Rate & 0.28 & 0.23 \\ 
$[111,48 ]$ & PENN World Table & 2.25 & 0.47 \\
   \hline
\end{tabular}
\end{table}

To see how these differences result in differently powered estimators, we perform another simulation exercise. For this exercise, we generate data using the variances estimated in Table \ref{tab:df_chisq_transformed} along with the true matrix sizes for the California Smoking,  the German reunification, and the Mariel boatlift data, following the procedure in Algorithm \ref{alg:real_data_simulation}. We use this data to compute and produce the resulting power curves in Figure \ref{fig:realistic_sim_studies_sc} below.

\begin{algorithm}[H]
\SetAlgoLined
\KwData{Empirical degrees of freedom $\hat{K}_{\nu}=2/\hat{\mmv}(\exp(\hat\nu_i))$, $\hat{K}_{\xi}=2/\hat{\mmv}(\exp(\hat\xi_t))$, simulation parameters $R$, $N$, $T$}
\KwResult{Power curves comparing marginal, unit placebo, time placebo, and conditional variance estimators}

\For{simulation $r = 1$ to $R$}{
   Draw unit multipliers $\exp(\nu_i) \sim \chi^2(\hat{K}_{\nu})/\hat{K}_{\nu}$ for $i = 1, \ldots, N$\;
   Draw time multipliers $\exp(\xi_t) \sim \chi^2(\hat{K}_{\xi})/\hat{K}_{\xi}$ for $t = 1, \ldots, T$\;
   
   Create variance matrix $\sigma^2_{it} = \exp(\nu_i)_i \cdot \exp(\xi_t)_t$\;
   
   \For{$i = 1$ to $N$, $t = 1$ to $T$}{
      Draw $Y_{it} \sim \mathcal{N}(0, \sigma^2_{it})$\;
   }
   
   Normalize $Y^{(r)} = \frac{Y^{(r)} - \bar{Y}^{(r)}}{sd(Y^{(r)})}$\;
}

\caption{Generate Real Data Simulation with empirical variances}\label{alg:real_data_simulation}
\end{algorithm}

In Figure \ref{fig:realistic_sim_studies_sc} we find that the C and the TP estimators perform similarly when $\hat{\mmv}(\exp(\hat\xi_t))$ is small, as in the California Smoking or Mariel Boatlift data. In contrast, when $\hat{\mmv}(\exp(\hat\nu_i))$ and $\hat{\mmv}(\exp(\hat\xi_t))$ are greater than 1--indicating that both heteroskedasticity over unit and time are important--the C estimator outperforms all of the other estimators. This is illustrated by the Germany Unification data where $\hat{\mmv}(\exp(\hat\nu_i)) = 2.14$ and $\hat{\mmv}(\exp(\hat\xi_t))=1.30$. Similar results hold for SDID (Appendix \ref{fig:realistic_sim_studies_sdid}) and TWFE (Appendix \ref{fig:realistic_sim_studies_twfe}) estimators.

\begin{figure}[H]
    \centering
        \caption{Power Curves: Semi-Synthetic Simulations Using SC Estimator}
    \begin{subfigure}{0.5\textwidth}
        \centering
        \includegraphics[width=\textwidth]{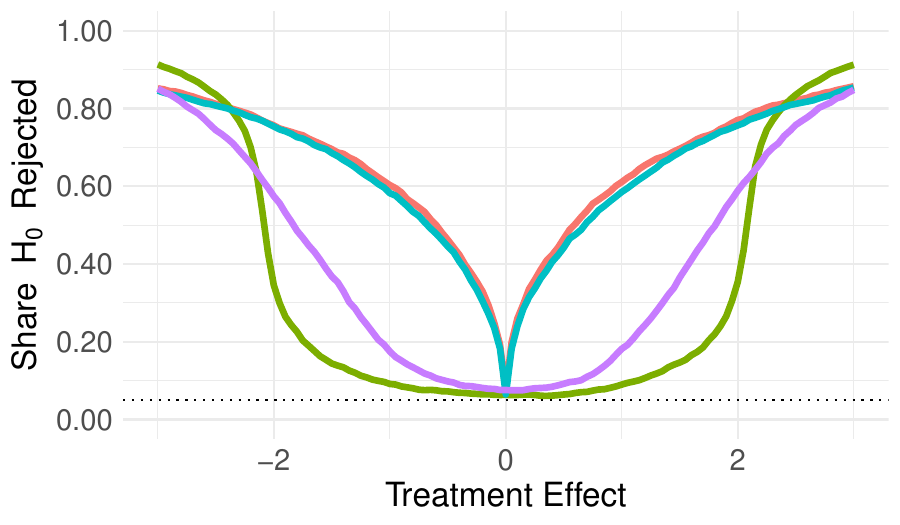}
        \caption{California Smoking: \\ ${\hat\mmv}(\exp(\hat\nu_i))=4.64$ and ${\hat\mmv}(\exp(\hat\xi_t))= 0.36$}
    \end{subfigure}
    \begin{subfigure}{0.5\textwidth}
        \centering
        \includegraphics[width=\textwidth]{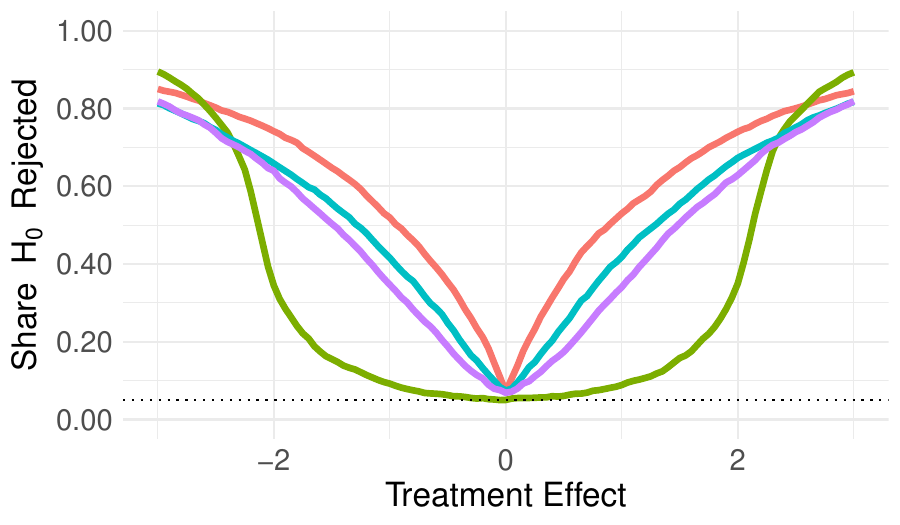}
        \caption{Germany Unification: \\ ${\hat\mmv}(\exp(\hat\nu_i))=2.14$ and ${\hat\mmv}(\exp(\hat\xi_t))= 1.30$}
    \end{subfigure}
    \begin{subfigure}{0.5\textwidth}
        \centering
        \includegraphics[width=1.1\textwidth]{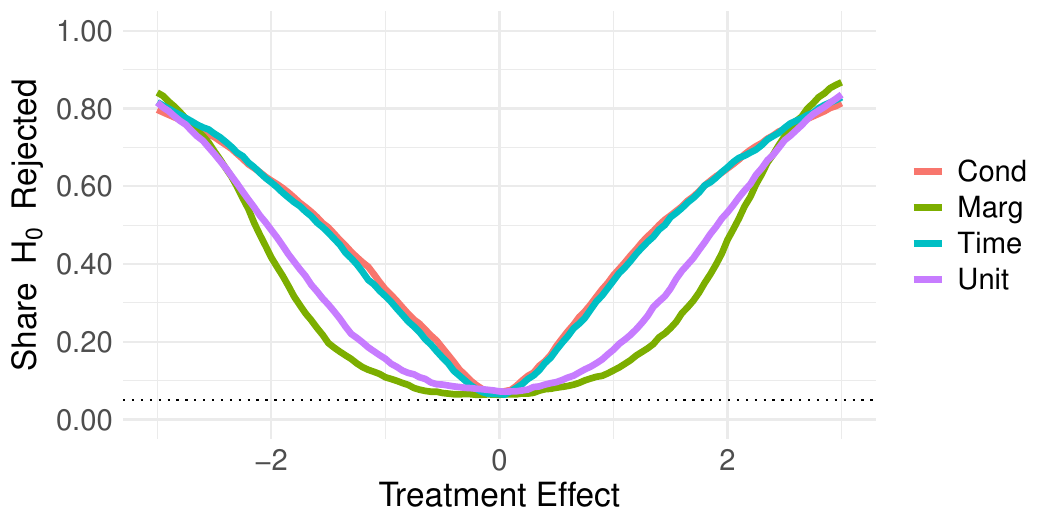}
        \caption{Mariel Boatlift\\ ${\hat\mmv}(\exp(\hat\nu_i))=1.26$ and ${\hat\mmv}(\exp(\hat\xi_t)) = 0.19$}
    \end{subfigure}
    \label{fig:realistic_sim_studies_sc}
\end{figure}

These findings are consistent with the fully simulated results in Figure \ref{fig:simple_sim_studies_sc}. At the same time, they highlight that real datasets can exhibit substantial heteroskedasticity in both the unit and time dimensions. As a result, it is clear that the choice between different variance estimators could be consequential for the power of empirical analysis. 

\section{Extending to multi-unit/period setting}

For expositional clarity, we have focused on the case with a single treated unit/time period pair $(\istar,\tstar)$. However, the setting with multiple treated units and/or multiple treated time periods is common, and both our framework and conditional estimator naturally extend to this case. Here we propose a simple way of dealing with treated block assignment for sufficiently large matrices, with modest fractions of treated unit/period pairs. 
Further research is needed to extend this framework to any complex assignment matrix, including those with units switching in and out of treatment. 

In the case with a single treated cell that was the focus of the earlier sections we estimated the variance by cycling over single placebo cells.
Under block assignment we cycle instead over placebo treatment blocks that match the treated block’s dimensions ($N_1$ units and $T_1$ time periods). For each candidate unit $j$ and start time $s$, we form a placebo block of size $N_1 \times T_1$, estimate any previously defined $g(\cdot)$ for that pseudo-block, and place the resulting residual in the residual matrix at position $(j,s)$. 
We drop candidate unit/time pairs $(i,s)$ for which it is not feasible to construct such placebo blocks. After cycling through all relevant blocks, we can compute the M, UP, TP, or C estimators using these residuals, now indexed by block rather than cell. This approach places greater demands on the size of the outcome matrix $\by$: the treated block must be small relative to $\mathbf{Y}$, or few valid placebo blocks will exist and the resulting variance estimators will be of poor quality. We provide further details in Appendix \ref{appx:block}.

\section{Conclusion}

In this paper we make two contributions. First, we develop a unified framework for understanding variance estimation in causal panel data models, encompassing several variance estimators that have previously been proposed for panel data based on non-nested sets of assumptions. Within this framework, we clarify the relative merits of these variance estimators. We show that previously proposed M, UP, and TP variance estimators can each be interpreted as placebo estimators targeting different (conditional) variances, and illustrate that these estimators can produce strikingly different standard errors. We find that these differences are not necessarily a matter of validity--under some set of assumptions all three are valid--but rather a matter of power, and that these power differences are driven by the extent and form of heteroskedasticity in the data.

Second, building on these insights, we propose a new Conditional (C) variance estimator which flexibly accounts for heteroskedasticity across both units and time. In realistic simulation settings, we find that this C estimator results in improved power for hypothesis testing. 

Our results underscore that the choice of variance estimator is an important decision. In making this decision researchers should take into account the heteroskedasticity structure of their data, and flexible approaches such as our proposed  C estimator are a practical and powerful new option.

\newpage
\bibliography{\bib}

\newpage
\appendix
\section*{Appendices}
\addcontentsline{toc}{section}{Appendices}
\renewcommand{\thesubsection}{\Alph{subsection}}

\subsection{Variance Estimation for Simulations}
\label{appendix:var}

Here we describe in detail the choice of
$\hat{\mmv}(\nu_i)$ in the simulations.
We first standardize the residuals so they have standard deviation one.
We estimate a two-way-fixed-effect model on the log of the squared standardized  residuals,  
\[ \ln(\hat \varepsilon_{it}^2)=\nu_i+\xi_t.\]
The estimated  $\exp(\tilde\nu_i)$ and $\exp(\tilde\xi_t)$ are rescaled to have mean one
\[ \hat\nu_i= \ln\left(\frac{\exp(\tilde\nu_i)}{\frac{1}{N}\sum_{j=1}^N \exp(\tilde\nu_j)}\right)\quad{\rm and}
\qquad
\hat\xi_t=\ln\left( \frac{\exp(\tilde\xi_t)}{\frac{1}{N}\sum_{s=1}^T \exp(\tilde\xi_s)}\right)
.\]
Next we calculate the variance of  the $\exp(\hat\nu_i)$:
\begin{align*}
    \hat{\mmv}(\exp(\hat\nu_i))&=\frac{1}{N}\sum_{i=1}^N\left(\exp(\hat\nu_i)-\frac{1}{N}\sum_{j=1}^N \exp(\hat\nu_j)\right)^2 .
\end{align*}
We model the $\exp(\nu_i)$ as draws from a chi-squared distribution with degrees of freedom equal to $K_\nu$ divided by $K_\nu$ to make it mean 1 and its variance is $2K_\nu/K^2_\nu=2/K_\nu$. Hence,
\[ 
K_\nu=\frac{2}{\hat{\mmv}(\exp(\hat\nu_i)))}\]
and similarly for $K_\xi$.

\clearpage
\newpage 

\subsection{Tables}
\setcounter{table}{0}
\renewcommand{\thetable}{B\arabic{table}}

\begin{table}[h!]
    \centering
    \caption{Comparison of Standard Errors for Two-way Fixed Effet Estimator Across Datasets}
    \centering
    \resizebox{\ifdim\width>\linewidth\linewidth\else\width\fi}{!}{
    \begin{tabular}[t]{lcccccc}
    \toprule
    \multicolumn{3}{c}{ } & \multicolumn{4}{c}{Standard Errors} \\
            Dataset & Treated 
        &  Point & Marginal & Unit & Time & Conditional \\
        &Period& Estimate&& Placebo & Placebo\\
        &  & & (M) & (UP) & (TP) & (C) \\
        \midrule
    California Smoking & 1989 & -12.904 & 0.367 & 0.510 & 0.271 & 0.374\\
    German Unification & 1990 & 1.960 & 0.046 & 0.098 & 0.032 & 0.151\\
    Mariel Boatlift & 1980 & 0.048 & 0.008 & 0.006 & 0.005 & 0.003\\
    \bottomrule
    \end{tabular}}
    \label{tab_app:table1_twfe}
\end{table}

\begin{table}[h!]
    \centering    
    \caption{Comparison of Standard Errors for Two-way Fixed Effet Estimator Across Datasets}
    \centering
    \resizebox{\ifdim\width>\linewidth\linewidth\else\width\fi}{!}{
    \begin{tabular}[t]{lcccccc}
    \toprule
        \toprule
        \multicolumn{3}{c}{ } & \multicolumn{4}{c}{Standard Errors} \\
        Dataset & Treated 
        &  Point & Marginal & Unit & Time & Conditional \\
        &Period& Estimate&& Placebo & Placebo\\
        &  & & (M) & (UP) & (TP) & (C) \\
        \midrule
        California Smoking & 1989 & -4.168 & 0.124 & 0.134 & 0.073 & 0.092\\
        German Unification & 1990 & 0.321 & 0.006 & 0.017 & 0.002 & 0.005\\
        Mariel Boatlift & 1980 & 0.099 & 0.009 & 0.005 & 0.005 & 0.001\\
        \bottomrule
        \end{tabular}}
    \label{tab_app:table1_sdid}
\end{table}

\begin{table}[H]
\centering
\caption{Standard Errors for the SDID Estimator}
\begin{tabular}{lccccc}
\hline
&& \multicolumn{4}{c}{Average Standard Errors}
\\
&& \multicolumn{4}{c}{(standard deviation)}
\\
Data Set & S.D.($\hat{\tau}_{\istar\tstar}$)  & \multicolumn{1}{c}{UP}
& \multicolumn{1}{c}{TP}
& \multicolumn{1}{c}{M}
& \multicolumn{1}{c}{C}\\
\hline
California Smoking & 0.58 & 0.55 & 0.51 & 0.60 & 0.52 \\
& & (0.24) & (0.30) & (0.04) & (0.40) \\
Germany & 1.05 & 0.86 & 0.84 & 1.05 & 0.80 \\
& & (0.64) & (0.64) & (0.08) & (0.94) \\
Mariel Boat & 1.17 & 1.02 & 0.76 & 1.14 & 0.88 \\
& & (0.47) & (0.88) & (0.14) & (1.17) \\
CPS hours & 1.06 & 0.99 & 0.96 & 1.07 & 0.93 \\
& & (0.39) & (0.47) & (0.03) & (0.62) \\
CPS log wage & 0.94 & 0.86 & 0.87 & 0.93 & 0.84 \\
& & (0.33) & (0.35) & (0.02) & (0.50) \\
CPS urate & 0.76 & 0.71 & 0.70 & 0.75 & 0.71 \\
& & (0.22) & (0.27) & (0.02) & (0.38) \\
PENN & 0.34 & 0.34 & 0.33 & 0.37 & 0.32 \\
& & (0.11) & (0.16) & (0.02) & (0.19) \\
\hline
\end{tabular}
\label{table:variance_comparison_table_sdid_new_format}
\begin{tablenotes}
\small
\item \textit{Note: This table reports  standard errors based on the four variance estimators, with their standard deviations in parentheses. The column S.D.($\hat{\tau}$) reports the standard deviation of the point estimate over the repeated samples.} 
\end{tablenotes}
\end{table}

\begin{table}[H]

\centering
\caption{Standard Errors for the TWFE Estimator}
\begin{tabular}{lccccc}
\hline
&& \multicolumn{4}{c}{Average Standard Errors}
\\
&& \multicolumn{4}{c}{(standard deviation)}
\\
Data Set & S.D.($\hat{\tau}_{\istar\tstar}$)  & \multicolumn{1}{c}{UP}
& \multicolumn{1}{c}{TP}
& \multicolumn{1}{c}{M}
& \multicolumn{1}{c}{C}\\
\hline
California Smoking & 0.69 & 0.63 & 0.59 & 0.67 & 0.61 \\
& & (0.23) & (0.33) & (0.03) & (0.45) \\
Germany & 1.04 & 0.85 & 0.84 & 1.02 & 0.79 \\
& & (0.57) & (0.60) & (0.02) & (0.87) \\
Mariel Boat & 1.10 & 0.98 & 0.73 & 1.06 & 0.83 \\
& & (0.40) & (0.79) & (0.07) & (0.98) \\
CPS hours & 0.99 & 0.91 & 0.90 & 0.99 & 0.85 \\
& & (0.37) & (0.44) & (0.01) & (0.55) \\
CPS log wage & 0.86 & 0.80 & 0.80 & 0.86 & 0.78 \\
& & (0.31) & (0.31) & (0.01) & (0.46) \\
CPS urate & 0.77 & 0.74 & 0.74 & 0.78 & 0.75 \\
& & (0.21) & (0.23) & (0.01) & (0.35) \\
PENN & 0.61 & 0.60 & 0.56 & 0.61 & 0.59 \\
& & (0.15) & (0.26) & (0.02) & (0.33) \\
\hline
\end{tabular}
\label{table:variance_comparison_table_twfe_new_format}
\begin{tablenotes}
\small
\item \textit{Note: This table reports standard errors based on the four variance estimators, with their standard deviations in parentheses. The column S.D.($\hat{\tau}$) reports the standard deviation of the point estimate over the repeated samples.} 
\end{tablenotes}
\end{table}

\begin{table}[H]
\centering
\caption{Coverage Rates for Nominally 95\% Confidence Intervals for the SDID Estimator}
\begin{tabular}{p{6cm}cccc}
\hline
Data Set & \multicolumn{1}{c}{UP}
& \multicolumn{1}{c}{TP}
& \multicolumn{1}{c}{M}
& \multicolumn{1}{c}{C}\\
\hline
California Smoking & 0.93 & 0.92 & 0.95 & 0.93 \\
Germany & 0.89 & 0.92 & 0.94 & 0.92 \\
Mariel Boat & 0.92 & 0.87 & 0.94 & 0.88 \\
CPS hours & 0.93 & 0.93 & 0.94 & 0.93 \\
CPS log wage & 0.92 & 0.93 & 0.93 & 0.92 \\
CPS urate & 0.93 & 0.93 & 0.94 & 0.93 \\
NA & 0.93 & 0.94 & 0.95 & 0.93 \\
\hline
\end{tabular}
\label{table:coverage_comparison_table_sdid}
\end{table}
\begin{table}[H]
\centering
\caption{Coverage Rates for Nominally 95\% Confidence Intervals for the TWFE Estimator}
\begin{tabular}{p{6cm}cccc}
\hline
Data Set & \multicolumn{1}{c}{UP}
& \multicolumn{1}{c}{TP}
& \multicolumn{1}{c}{M}
& \multicolumn{1}{c}{C}\\
\hline
California Smoking & 0.93 & 0.92 & 0.94 & 0.93 \\
Germany & 0.90 & 0.92 & 0.94 & 0.93 \\
Mariel Boat & 0.93 & 0.88 & 0.95 & 0.90 \\
CPS hours & 0.93 & 0.94 & 0.94 & 0.93 \\
CPS log wage & 0.94 & 0.92 & 0.94 & 0.93 \\
CPS urate & 0.94 & 0.94 & 0.95 & 0.94 \\
NA & 0.94 & 0.95 & 0.94 & 0.95 \\
\hline
\end{tabular}
\label{table:coverage_comparison_table_twfe}
\end{table}

\subsection{Graphs}
\setcounter{figure}{0}
\renewcommand{\thefigure}{C\arabic{figure}}

\begin{figure}[htbp!]
    \centering
    \caption{Power Curves: Illustrative Simulations Using SDID Estimator.}        
     \begin{subfigure}{0.48\textwidth}
        \centering
        \includegraphics[width=\textwidth]{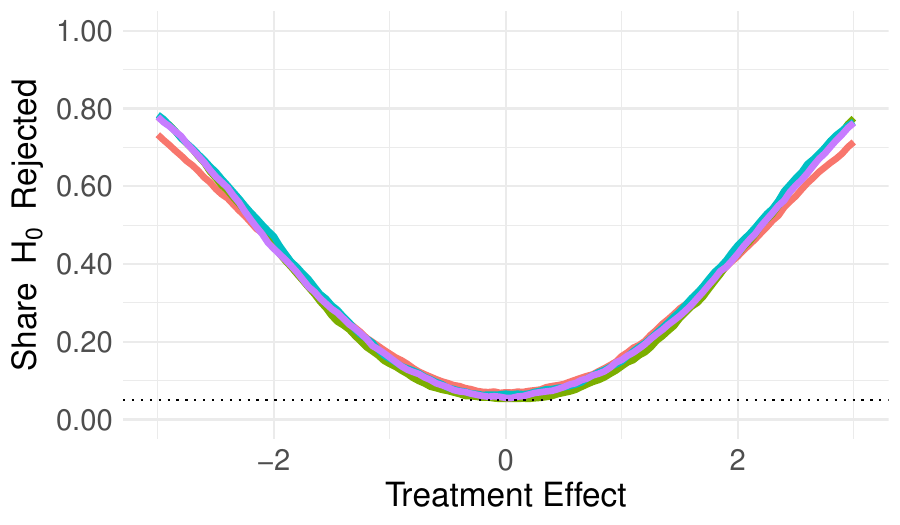}
        \caption{$\mmv(\exp(\nu_i))=\mmv(\exp(\xi_t))=0$}
    \end{subfigure}
\begin{subfigure}{0.48\textwidth}
        \centering
        \includegraphics[width=\textwidth]{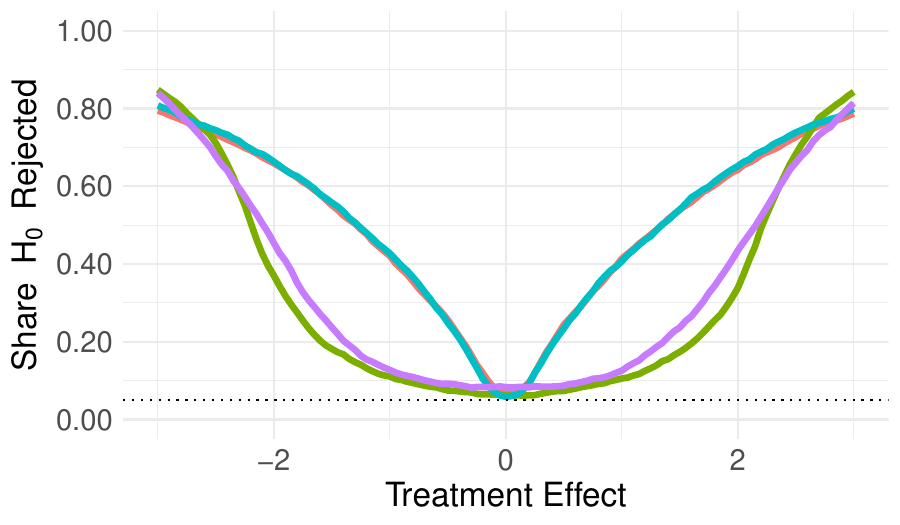}
        \caption{$\mmv(\exp(\nu_i))=2$, $\mmv(\exp(\xi_t))=0$}
    \end{subfigure}

 \begin{subfigure}{0.48\textwidth}
        \centering
        \includegraphics[width=\textwidth]{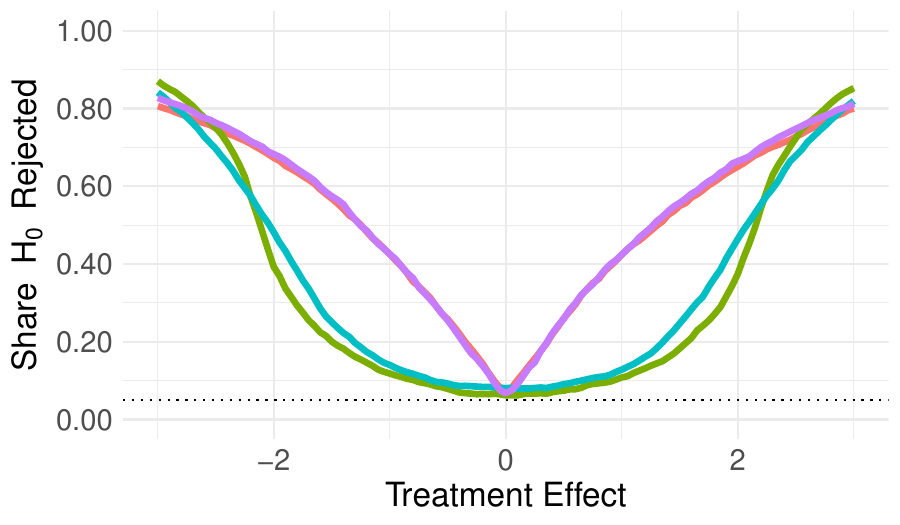}
        \caption{$\mmv(\exp(\nu_i))=0$, $\mmv(\exp(\xi_t))=2$}
    \end{subfigure}
     \begin{subfigure}{0.48\textwidth}
        \centering
        \includegraphics[width=1.2\textwidth]{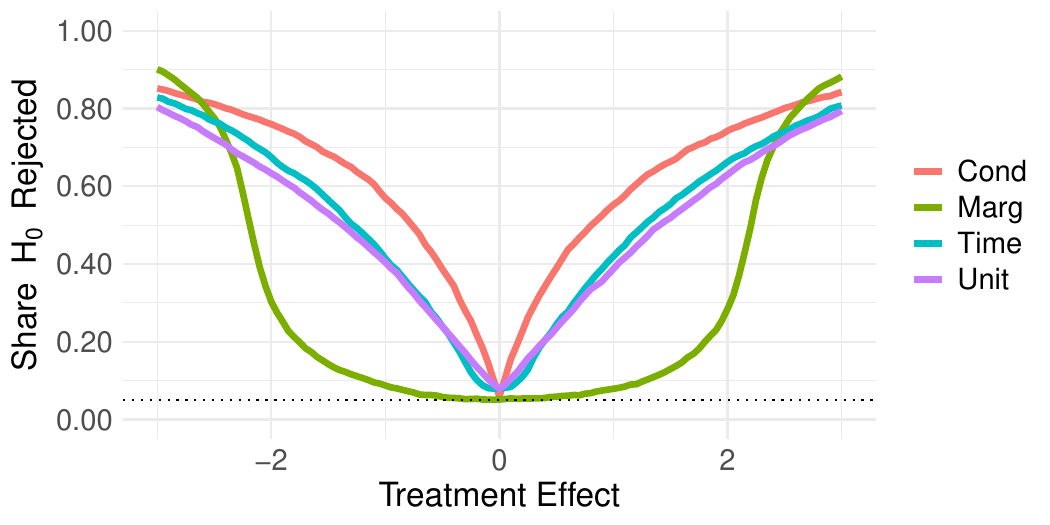}
        \caption{$\mmv(\exp(\nu_i))=\mmv(\exp(\xi_t))=2$}
    \end{subfigure}
    \label{fig:simple_sim_studies_sdid}
\end{figure}

\begin{figure}[htbp!]
    \centering
    \caption{Power Curves: Illustrative Simulations Using TWFE Estimator.}        
     \begin{subfigure}{0.48\textwidth}
        \centering
        \includegraphics[width=\textwidth]{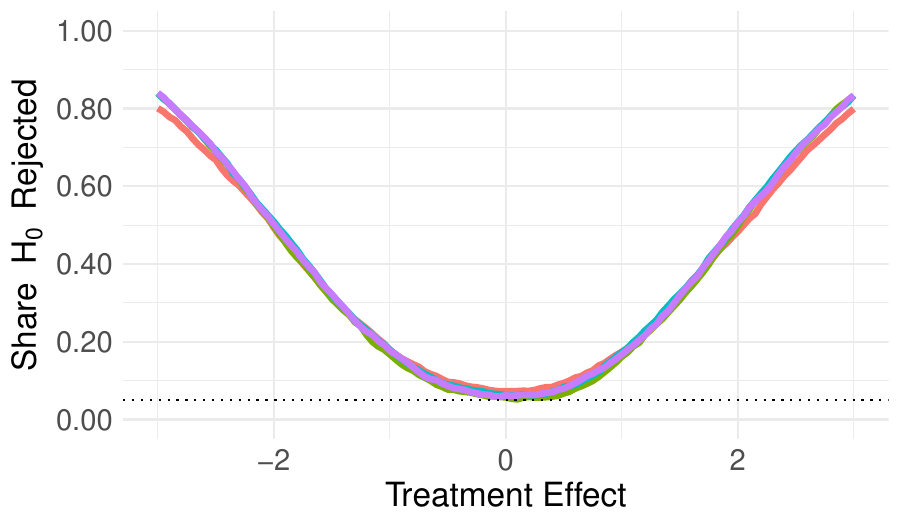}
        \caption{$\mmv(\exp(\nu_i))=\mmv(\exp(\xi_t))=0$}
    \end{subfigure}
\begin{subfigure}{0.48\textwidth}
        \centering
        \includegraphics[width=\textwidth]{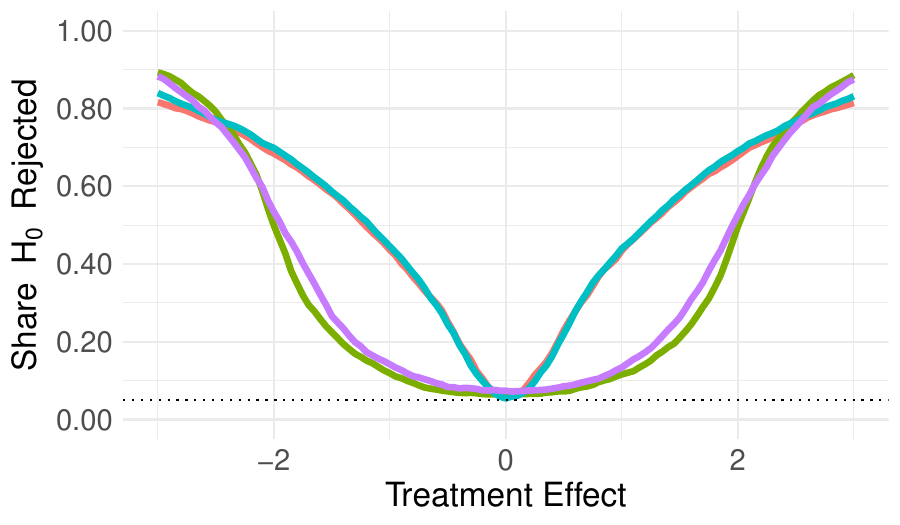}
        \caption{$\mmv(\exp(\nu_i))=2$, $\mmv(\exp(\xi_t))=0$}
    \end{subfigure}

 \begin{subfigure}{0.48\textwidth}
        \centering
        \includegraphics[width=\textwidth]{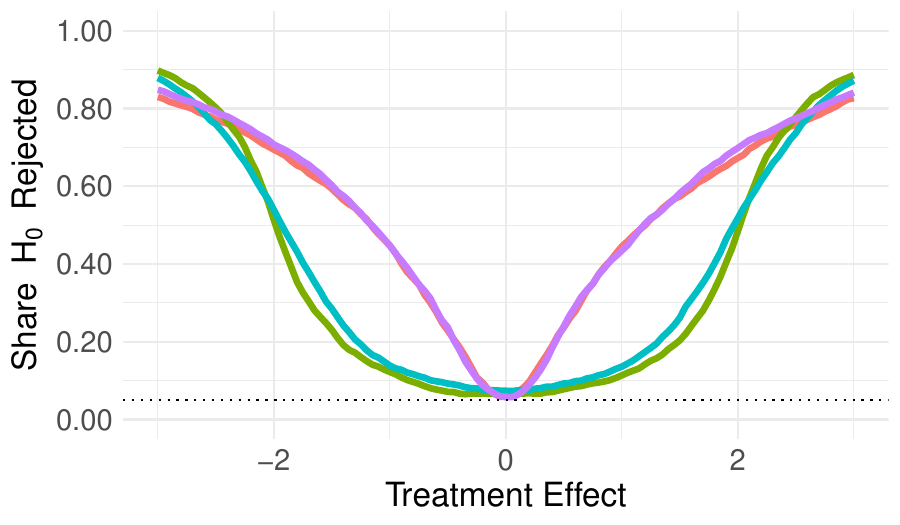}
        \caption{$\mmv(\exp(\nu_i))=0$, $\mmv(\exp(\xi_t))=2$}
    \end{subfigure}
     \begin{subfigure}{0.48\textwidth}
        \centering
        \includegraphics[width=1.2\textwidth]{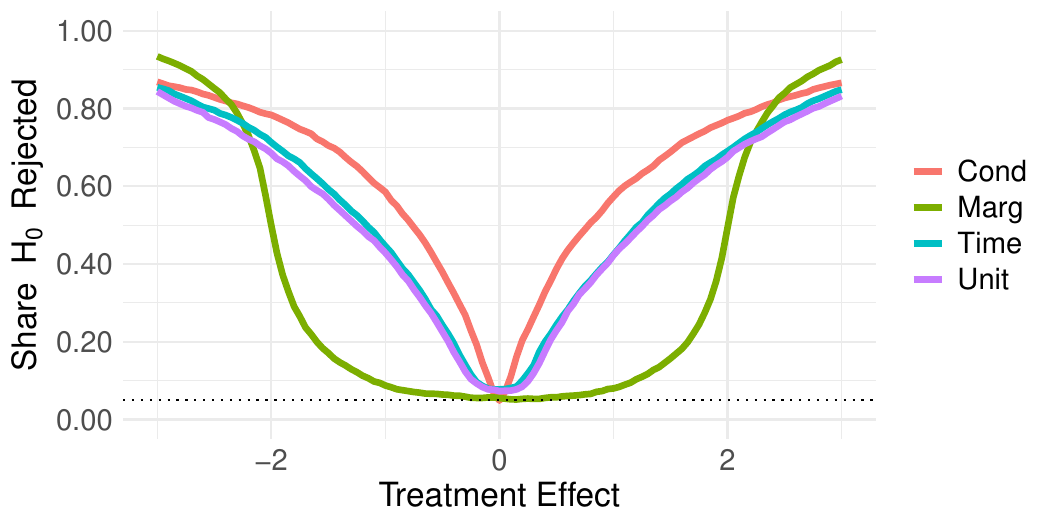}
        \caption{$\mmv(\exp(\nu_i))=\mmv(\exp(\xi_t))=2$}
    \end{subfigure}
    \label{fig:simple_sim_studies_twfe}
\end{figure}

\begin{figure}[H]
    \centering
        \caption{Power Curves: Semi-synthetic Simulations using SDID estimator}
    \begin{subfigure}{0.5\textwidth}
        \centering
        \includegraphics[width=\textwidth]{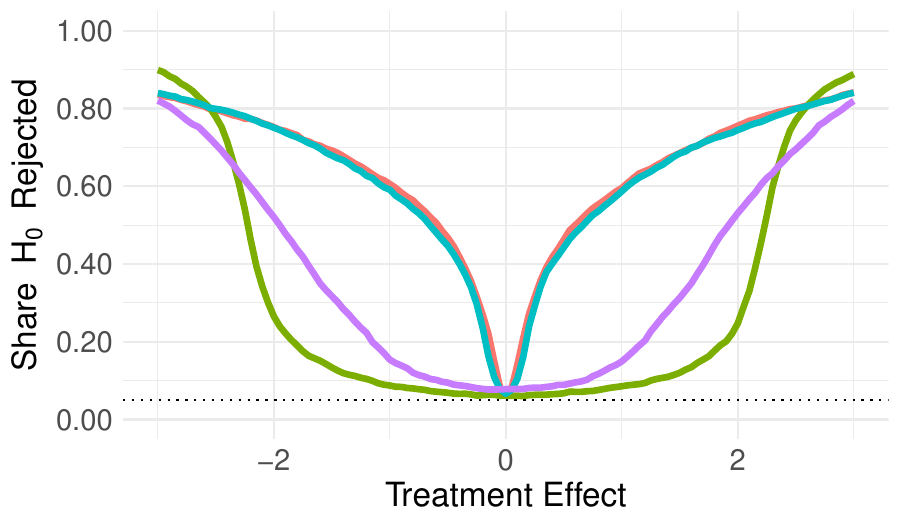}
        \caption{California Smoking: \\ ${\mmv}(\exp(\hat\nu_i))=4.64$ and ${\mmv}(\exp(\hat\xi_t))= 0.36$}
    \end{subfigure}
    \begin{subfigure}{0.5\textwidth}
        \centering
        \includegraphics[width=\textwidth]{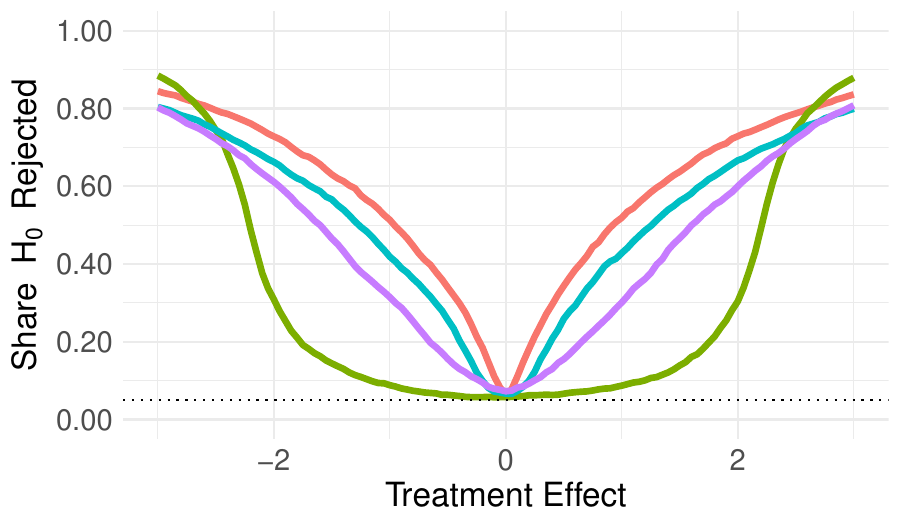}
        \caption{Germany Unification: \\ ${\mmv}(\exp(\hat\nu_i))=2.14$ and ${\mmv}(\exp(\hat\xi_t))= 1.30$}
    \end{subfigure}
    \begin{subfigure}{0.5\textwidth}
        \centering
        \includegraphics[width=1.1\textwidth]{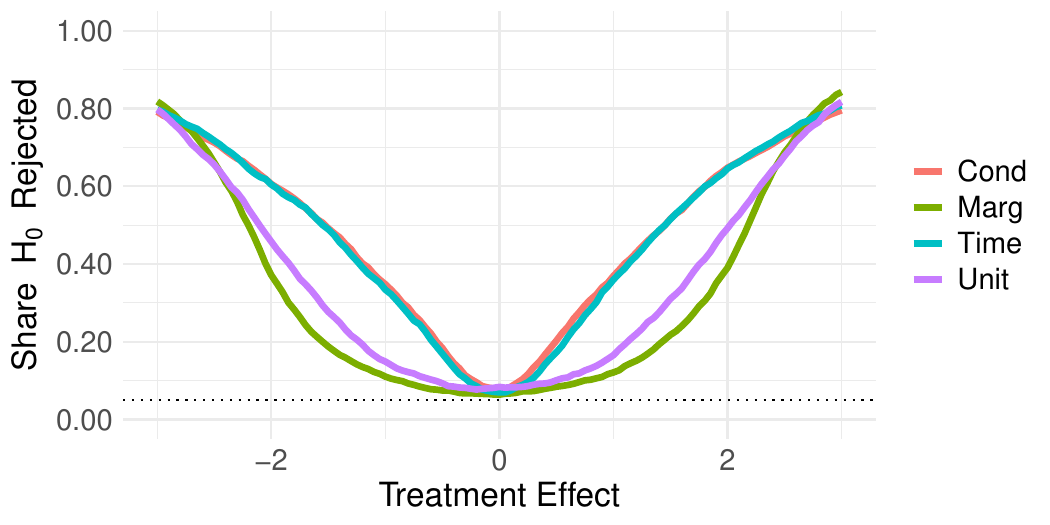}
        \caption{Mariel Boatlift\\ ${\mmv}(\exp(\hat\nu_i))=1.26$ and ${\mmv}(\exp(\hat\xi_t)) = 0.19$}
    \end{subfigure}%
    \label{fig:realistic_sim_studies_sdid}
\end{figure}

\begin{figure}[H]
    \centering
        \caption{Power Curves: Semi-synthetic Simulations using TWFE estimator}
    \begin{subfigure}{0.5\textwidth}
        \centering
        \includegraphics[width=\textwidth]{_images/_main_paper/simulation_sdid_2500_50_40_cal_Y_df.csv.pdf}
        \caption{California Smoking: \\ $\hat{\mmv}(\exp(\hat\nu_i))=4.64$ and $\hat{\mmv}(\exp(\hat\xi_t))= 0.36$}
    \end{subfigure}%
    \begin{subfigure}{0.5\textwidth}
        \centering
        \includegraphics[width=\textwidth]{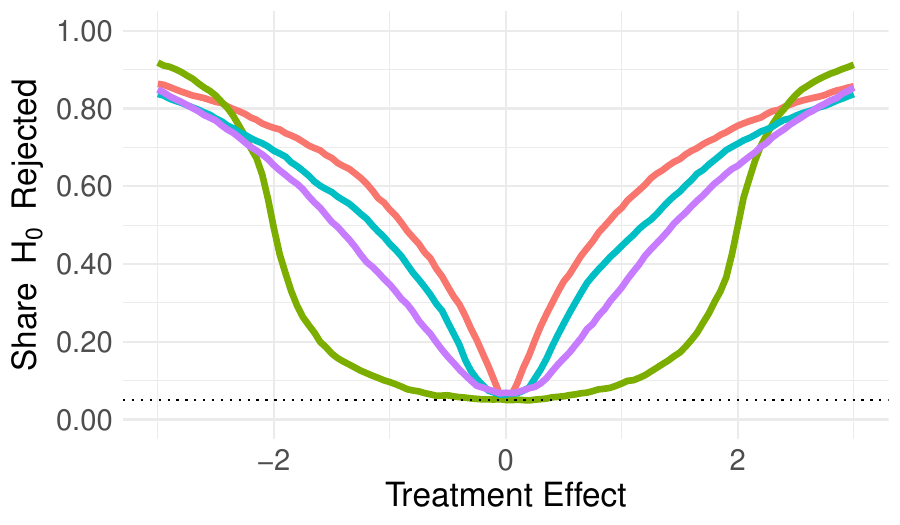}
        \caption{Germany Unification: \\ $\hat{\mmv}(\exp(\hat\nu_i))=2.14$ and $\hat{\mmv}(\exp(\hat\xi_t))= 1.30$}
    \end{subfigure}
    \begin{subfigure}{0.5\textwidth}
        \centering
        \includegraphics[width=1.1\textwidth]{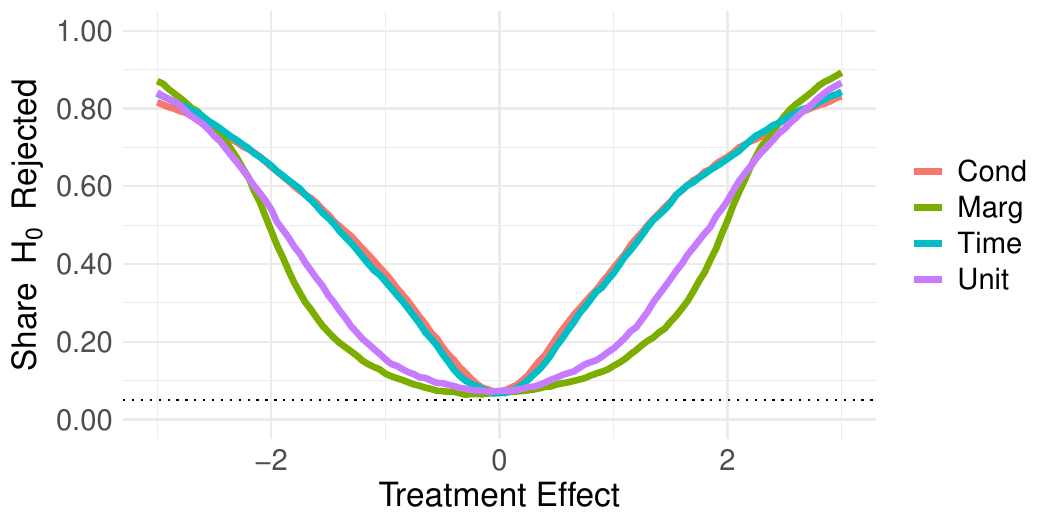}
        \caption{Mariel Boatlift\\ $\hat{\mmv}(\exp(\hat\nu_i))=1.26$ and $\hat{\mmv}(\exp(\hat\xi_t)) = 0.19$}
    \end{subfigure}%
    \label{fig:realistic_sim_studies_twfe}
\end{figure}

\subsection{Proofs}\label{appx:proofs}

\paragraph{\textbf{Proof of Proposition} \ref{prop1}}
First consider $(i)$. By Assumption \ref{assumption:exp}
$\mu(\kappa,\nu,\xi)=0$, so $\mu(\kappa,\nu)=\mathbb{E}[\mu(\kappa,\nu,\xi)]=0$, and teh same applies for $\mu(\kappa,\xi)$ and $\mu(\kappa).$

Next, consider $(ii)$. First, the variance of $\varepsilon_{it}$ is by definition,
\[ \mmv(\varepsilon_{it})=\mme[(\varepsilon_{it}-\mme[\varepsilon_{it}])^2
=\mme[\varepsilon^2_{it}]
=\sigma^2(\kappa).\]
Next, 
\[\sigma^2(\kappa,\nu_i)=\mathbb{E}[(\varepsilon^*_{it}-\mu(\kappa,\nu_i))^2|\kappa,\nu_i]\]
which by Assumption \ref{assumption:exp} is equal to
\[\mathbb{E}[\varepsilon^2_{it}|\kappa,\nu_i].\]
Taking the expectation over this leads to
\[ \mme[\sigma^2(\kappa,\nu_i)]
=\mathbb{E}[\mathbb{E}[\varepsilon^2_{it}|\kappa,\nu_i]]
=\mme[\varepsilon_{it}^2]=\mmv(\varepsilon_{it}).\]
The same argument works for the other parts of $(ii)$.

For part $(iii)$, consider
\[\sigma^2(\kappa) =\mme[(\sigma^2(\kappa,\nu))^2]
=\mmv(\sigma^2(\kappa,\nu)|\kappa)+
\mme[\sigma^2(\kappa,\nu)|\kappa]^2\]
implying that
\[\mmv(\sigma^{2}(\kappa)) \leq\mmv(\sigma^{2}(\kappa,\nu_i)),\]
with a similar argument for the other inequalities.
$\square$

\paragraph{\textbf{Proof of Proposition} \ref{prop2}}
Consider $\hat \sigma^{2,\marg}_{it}$:
\[ \hat \sigma^{2,\marg}_{it}=\frac{1}{NT-1}\sum_{(j,s)\neq (i,t)} \varepsilon_{js}^2
=\frac{1}{NT-1}\sum_{(j,s)\neq (i,t)} g(\kappa,\nu_j,\xi_t,\eta_{js})^2
\]
\[ = \frac{1}{NT}\sum_{(j,s)} g(\kappa,\nu_j,\xi_t,\eta_{js})^2+
\frac{1}{NT(NT-1)}\sum_{(j,s)\neq (i,t)} g(\kappa,\nu_j,\xi_t,\eta_{js})^2-
 \frac{1}{NT} g(\kappa,\nu_i,\xi_t,\eta_{it})^2
\]
The first term is
\[ \frac{1}{NT}\sum_{(j,s)} g(\kappa,\nu_j,\xi_t,\eta_{js})^2=\sigma^2(\kappa)+o_p().\]
The second and third terms are both $o_p(1)$. 

The other claims follow the same argument.
$\square$

\subsection{Algorithms}
\setcounter{algocf}{0}
\renewcommand{\thealgocf}{E\arabic{algocf}}
{
\begin{algorithm}[H]
\SetAlgoLined
\KwData{Panel data $\by$, unit $i$, time $t$, treated unit $\istar$, treated time $\tstar$}
\KwResult{Imputed counterfactual $\widehat{Y_{it}}(0)$}
Exclude observation $(i,t)$ and treated observation $(\istar,\tstar)$ from $\by$\;
Fit linear model: $Y_{it} = \alpha_i + \beta_t + \epsilon_{it}$\;
\Indp
where $\alpha_i$ are unit fixed effects\;
and $\beta_t$ are time fixed effects\;
\Indm
Compute counterfactual prediction: $\widehat{Y_{it}}(0) = \widehat{\alpha_i} + \widehat{\beta_t}$\;
\Return{counterfactual prediction $\widehat{Y_{it}}(0)$}
\caption{TWFE Counterfactual Prediction}\label{alg:twfe_estimation_var}
\end{algorithm}
}

\vspace{1cm}

{
\begin{algorithm}[H]
\SetAlgoLined
\KwData{Panel data $\by$, treatment matrix $\bw$, unit $i$, time $t$, treated unit $\istar$, treated time $\tstar$}
\KwResult{Imputed counterfactual $\widehat{Y_{it}}(0)$}
Set pseudo treatment: $W_{it} = 1$\;
\If{$i \neq \istar$}{
    Remove row $\istar$ from $Y$ and $W$\;
}{
}
\If{$t \neq \tstar$}{
    Remove column $\tstar$ from $Y$ and $W$\;
}{
}
Calculate SC weights $\hat{\omega}$ using pre-treatment data\;
Compute counterfactual weighted combination: $\widehat{Y_{it}}(0) = Y_{it} - \sum_{k \neq i, \istar} \hat{\omega}_j Y_{jt}$\;
\Return{counterfactual weighted SC combination $\widehat{Y_{it}}(0)$}
\caption{SC Counterfactual Prediction}\label{alg:sc_estimation_var}
\end{algorithm}
}

\vspace{1cm}

{
\begin{algorithm}[H]
\SetAlgoLined
\KwData{Panel data $\by$, treatment matrix $\bw$, unit $i$, time $t$, treated unit $\istar$, treated time $\tstar$}
\KwResult{Imputed counterfactual $\widehat{Y_{it}}(0)$}
Set pseudo treatment: $W_{it} = 1$\;
\If{$i \neq \istar$}{
    Remove row $\istar$ from $Y$ and $W$\;
}{
}
\If{$t \neq \tstar$}{
    Remove column $\tstar$ from $Y$ and $W$\;
}{
}
Calculate unit weights $\hat{\omega}$ and time weights $\hat{\lambda}$\;
Compute counterfactual weighted combination: $\widehat{Y_{it}}(0) = Y_{it} - \sum_{(j,s) \neq (i,t), (\istar, \tstar)} \hat{\omega}_j \hat{\lambda}_s Y_{js}$\;
\Return{counterfactual weighted SDID combination $\widehat{Y_{js}}(0)$}
\caption{SDID Counterfactual Prediction}\label{alg:sdid_estimation_var}
\end{algorithm}
}

\vspace{1cm}

\textbf{Realistic data generation procedure}. In this section, we provide the details of the realistic data generation procedure. There are two main steps, (1) the DGP estimation and (2) the data simulation. This procedure broadly follows the procedure outlined in \cite{arkhangelsky2021synthetic}, with the addition of more explicitly modeled time and unit heteroskedasticity. \\

\vspace{0.5cm}

\begin{algorithm}[H] 
\SetAlgoLined
\KwData{Panel data $\by$ (N × T), Assignment vector $\mathbf{W}$, Rank parameter $r$}
\KwResult{DGP parameters: $\mathbf{F}$, $\mathbf{M}$, $\boldsymbol{\Sigma}$, $\boldsymbol{\pi}$, AR coefficients, unit\_var, time\_var}

\textbf{Step 1: Data Decomposition}\;
Normalize panel data: $\by \leftarrow (\by - \text{mean}(\by))/\text{sd}(\by)$\;
Decompose $\by$ into systematic and idiosyncratic components:\;
\quad $\by = \mathbf{L} + \mathbf{E} = (\mathbf{F} + \mathbf{M}) + \mathbf{E}$\;
\quad where systematic component $\mathbf{L} = \mathbf{U}\mathbf{D}\mathbf{V}^\top$ via SVD with rank $r$\;
\quad $\mathbf{F} \leftarrow$ additive fixed effects (row + column means of $\mathbf{L}$)\;
\quad $\mathbf{M} \leftarrow \mathbf{L} - \mathbf{F}$ (interactive effects)\;
\quad $\mathbf{E} \leftarrow \by - \mathbf{L}$ (residuals)\;
Store unit factors: $\mathbf{U}_r \leftarrow$ first $r$ columns of $\mathbf{U} \times \sqrt{N}$\;

\textbf{Step 2: Variance Component Estimation}\;
Fit variance model:\;
\quad $\log(\mathbf{E}_{it}^2) = \kappa + \gamma_i + \lambda_t + \nu_{it}$\;
Extract unit- and time-specific variance effects:\;
\quad unit\_var$_i \leftarrow \exp(\gamma_i)$ for $i = 1, \ldots, N$\;
\quad time\_var$_t \leftarrow \exp(\lambda_t)$ for $t = 1, \ldots, T$\;
\quad Normalize both time and unit effects

\textbf{Step 3: Autocorrelation Structure}\;
Fit AR(2) model to residuals:\;
\quad $E_{it} = \rho_1 E_{i,t-1} + \rho_2 E_{i,t-2} + \eta_{it}$\;
\quad ar\_coef $\leftarrow$ fit\_ar2($\mathbf{E}$)\;
\quad $\boldsymbol{\Sigma} \leftarrow$ ar2\_correlation\_matrix(ar\_coef, $T$)\;

\textbf{Step 5: Treatment Assignment Model}\;
Estimate treatment probability:\;
\quad $\boldsymbol{\pi} \leftarrow$ glm($\mathbf{W} \sim \mathbf{U}_r$, family = 'binomial')\$fitted.values\;
\quad where $\mathbf{U}_r \in \mathbb{R}^{N \times r}$ are the unit factors (first $r$ left singular vectors) from Step 1\;

\Return{$\mathbf{F}$, $\mathbf{M}$, $\boldsymbol{\Sigma}$, $\boldsymbol{\pi}$, ar\_coef, unit\_var, time\_var}

\caption{Realistic DGP: Parameter Estimation}\label{alg:dgp_estimation}
\end{algorithm}


\subsection{Block Placebo Variance for SC Estimator}\label{appx:block}

\noindent\textbf{Set-up.} 
For expositional clarity, we have focused on the case with a single treated unit–time period pair $(i^\ast,t^\ast)$. However, settings with multiple treated units and multiple treated time periods are common, and both our framework and conditional estimator naturally extend to this case. Under block assignment, for example, rather than cycling over single placebo cells, we cycle over placebo treatment {blocks} that match the treated block’s dimensions ($N_1$ units and $T_1$ time periods).

Let the treated pattern be a set of indices 
$\mathcal{T}_1 = \{(i,t): W_{it}=1\}$ 
of size $N_1 \times T_1$ ({\it e.g.}, $N_1$ units treated for $T_1$ consecutive periods starting at $T_0$). 
For each control unit $j$ and pre-treatment start time $s$ satisfying $s+T_1-1\le T_0$, define a placebo block
\[
\mathcal{B}(j,s) = \{(j,t): t\in\{s,\ldots,s+T_1-1\}\},
\]
and form a placebo treatment matrix $W^{(j,s)}$ that equals 1 on $\mathcal{B}(j,s)$ and 0 elsewhere. 
If there are $N_1>1$ simultaneously treated units, extend the domain of the first argument of $\mathcal{B}(j,s) $ to a unit set ${\cal U}\subset\{1,\ldots,N\}$ with $|{\cal U}|=N_1$, 
and define 
\[
\mathcal{B}({\cal U},s)=\{(j,t): j\in {\cal U},\ t\in\{s,\ldots,s+T_1-1\}\},
\]
with the corresponding placebo treatment matrix $W^{({\cal U},s)}$.

\medskip
\noindent\textbf{Illustrative example.}
To fix ideas, consider a toy panel with $N=5$ units and $T=6$ periods. 
Suppose the true treatment applies to units $4$–$5$ starting at $t=5$ and lasting $T_1=2$ periods, so $N_1=2$. 
Then the outcome and treatment matrices can be represented as

\[
Y =
\begin{bmatrix}
y_{11} & y_{12} & y_{13} & y_{14} & y_{15} & y_{16}\\
y_{21} & y_{22} & y_{23} & y_{24} & y_{25} & y_{26}\\
y_{31} & y_{32} & y_{33} & y_{34} & y_{35} & y_{36}\\
y_{41} & y_{42} & y_{43} & y_{44} & \mathbf{y_{45}} & \mathbf{y_{46}}\\
y_{51} & y_{52} & y_{53} & y_{54} & \mathbf{y_{55}} & \mathbf{y_{56}}
\end{bmatrix},
\qquad
W =
\begin{bmatrix}
0 & 0 & 0 & 0 & 0 & 0\\
0 & 0 & 0 & 0 & 0 & 0\\
0 & 0 & 0 & 0 & 0 & 0\\
0 & 0 & 0 & 0 & \mathbf{1} & \mathbf{1}\\
0 & 0 & 0 & 0 & \mathbf{1} & \mathbf{1}
\end{bmatrix}.
\]

Now consider a placebo block for units ${\cal U}=\{2,3\}$ starting at $s=3$. 
The corresponding placebo treatment matrix $W^{({\cal U},3)}$ is
\[
W^{({\cal U},3)} =
\begin{bmatrix}
0 & 0 & 0 & 0 & 0 & 0\\
0 & 0 & \mathbf{1} & \mathbf{1} & 0 & 0\\
0 & 0 & \mathbf{1} & \mathbf{1} & 0 & 0\\
0 & 0 & 0 & 0 & 1 & 1\\
0 & 0 & 0 & 0 & 1 & 1
\end{bmatrix}.
\]
Here, the placebo block $(U,s)$ mimics the true treated block’s shape ($N_1=2$, $T_1=2$) but occurs earlier in time. 
For each such block, we estimate a pseudo-effect $\widehat\tau_{U,s}^{\mathrm{SC}}$ using the SC estimator applied to the truncated panel up to period $s+T_1-1$.
Alternative placebo treated blocks are
\[
W^{({\cal U}',4)} =
\begin{bmatrix}
0 & 0 & 0 & \mathbf{1} & \mathbf{1} & 0\\
0 & 0 & 0 & \mathbf{1} & \mathbf{1} & 0\\
0 & 0 & 0 & 0 & 0 & 0\\
0 & 0 & 0 & 0 & 1 & 1\\
0 & 0 & 0 & 0 & 1 & 1
\end{bmatrix}\qquad {\rm or}\quad 
W^{({\cal U}'',3)} =
\begin{bmatrix}
0 & 0 & 0 & 0 & 0 & 0\\
0 & 0 & 0 & 0 & 0 & 0\\
0 & \mathbf{1} & \mathbf{1} & 0 & 0 & 0\\
0 & 0 & 0 & 0 & 1 & 1\\
0 & \mathbf{1} & \mathbf{1} & 0 & 1 & 1
\end{bmatrix}.
\]

\medskip
\noindent\textbf{Placebo effect for a block.}
For the the implementation we focus on the SC estimator.
Drop all periods after $s+T_1-1$, and all units who are treated in some periods between $t=0$ and $s+T_1-1$. 
Using Algorithm \ref{alg:sc_estimation_var}, fit unit weights on pre-block data to match the average of the placebo treated units and compute the average post-block pseudo-effect
\[
\widehat\tau_{{\cal U},s}^{\mathrm{SC}} 
= 
\frac{1}{N_1T_1}\sum_{j\in {\cal U}}\sum_{t=s}^{s+T_1-1}
\Big(Y_{jt}-\sum_{k\notin {\cal U}}\widehat\omega_{k}^{(U,s)}Y_{kt}\Big).
\]
Store this
indexed by the block’s first period $s$ and the set ${\cal U}$). 

\medskip
\noindent\textbf{Variance estimators as block averages.}
Replace the cell-wise residuals in M/UP/TP by block-level residuals and average squares over the corresponding placebo index sets:
\begin{align*}
\widehat\sigma^2_{\mathrm{M,block}} &= \text{mean of }(\widehat\tau_{\cdot,\cdot}^{\mathrm{SC}})^2\ \text{over all placebo blocks},\\
\widehat\sigma^2_{\mathrm{UP,block}} &= \text{mean over blocks aligned with the treated period(s)},\\
\widehat\sigma^2_{\mathrm{TP,block}} &= \text{mean over blocks aligned with the treated unit set}.
\end{align*}
This mirrors the single-cell case where M/UP/TP are averages of squared residuals over different index sets. Conceptually, we have coarsened the permutation unit from individual cells to blocks that preserve the treatment pattern.
As in the single-period case, reliable inference requires that the treated block be relatively small compared to the $\mathbf{Y}$ matrix, so that a sufficient number of placebo blocks can be formed. 
Extensions to more general treatment assignment patterns, including staggered adoption, remain an area for future work.

\end{document}